\documentclass[10pt,conference]{IEEEtran}
\usepackage{cite}
\usepackage{amsmath,amssymb,amsfonts}
\usepackage{algorithmic}
\usepackage{graphicx}
\usepackage{textcomp}
\usepackage{xcolor}
\usepackage[hyphens]{url}
\usepackage{fancyhdr}
\usepackage{hyperref}
\usepackage{inconsolata}
\usepackage{flushend}

\newcommand{\hpcayear}{2025}

\usepackage{pdfprivacy} 
\usepackage{xspace}      
\usepackage{graphicx}    
\usepackage{multirow}    
\usepackage{booktabs}    
\usepackage{caption}
\usepackage{algorithm}
\usepackage{algorithmic}
\usepackage{pifont}

\newcommand{\putsec}[2]{\section{#2}\label{sec:#1}}
\newcommand{\putssec}[2]{\subsection{#2}\label{ssec:#1}}

\newcommand{\secref}[1]{Section~\ref{sec:#1}}
\newcommand{\ssecref}[1]{Section~\ref{ssec:#1}}

\newcommand{\figref}[1]{Figure~\ref{fig:#1}}
\newcommand{\tabref}[1]{Table~\ref{tab:#1}}
\newcommand{\eqnref}[1]{Equation~\ref{eqn:#1}}
\newcommand{\algref}[1]{Algorithm~\ref{alg:#1}}

\newcommand{\name}{{VR-Pipe\xspace}}

\newcommand{\myparagraph}[1]{\vspace{0.02in}\noindent \textbf{#1}}

\title{\name{}: Streamlining Hardware Graphics Pipeline for Volume Rendering}

\def\hpcacameraready{} 

\newcommand\hpcaauthors{Junseo Lee \quad\quad Jaisung Kim \quad\quad Junyong Park \quad\quad Jaewoong Sim}
\newcommand\hpcaaffiliation{{} \\ {Seoul National University}}
\newcommand\hpcaemail{{\{junseo.lee, jaisung.kim, junyong.park, jaewoong\}@snu.ac.kr}}

\author{
  \ifdefined\hpcacameraready
    \IEEEauthorblockN{\hpcaauthors{}}
      \IEEEauthorblockA{
        \hpcaaffiliation{} \\
        \hpcaemail{}
      }
  \else
    \IEEEauthorblockN{\normalsize{HPCA \hpcayear{} Submission
      \textbf{\#\hpcasubmissionnumber{}}} \\
      \IEEEauthorblockA{
        Confidential Draft \\
        Do NOT Distribute!!
      }
    }
  \fi 
}

\fancypagestyle{camerareadyfirstpage}{%
  \fancyhead{}
  
  \fancyhead[C]{
    \ifdefined\aeopen
    \parbox[][12mm][t]{13.5cm}{\hpcayear{} IEEE International Symposium on High-Performance Computer Architecture (HPCA)}    
    \else
      \ifdefined\aereviewed
      \parbox[][12mm][t]{13.5cm}{\hpcayear{} IEEE International Symposium on High-Performance Computer Architecture (HPCA)}
      \else
      \ifdefined\aereproduced
      \parbox[][12mm][t]{13.5cm}{\hpcayear{} IEEE International Symposium on High-Performance Computer Architecture (HPCA)}
      \else
      \parbox[][0mm][t]{13.5cm}{\hpcayear{} IEEE International Symposium on High-Performance Computer Architecture (HPCA)}
    \fi 
    \fi 
    \fi 
    \ifdefined\aeopen 
      \includegraphics[width=12mm,height=12mm]{ae-badges/open-research-objects.pdf}
    \fi 
    \ifdefined\aereviewed
      \includegraphics[width=12mm,height=12mm]{ae-badges/research-objects-reviewed.pdf}
    \fi 
    \ifdefined\aereproduced
      \includegraphics[width=12mm,height=12mm]{ae-badges/results-reproduced.pdf}
    \fi
  }
  \fancyfoot[C]{}
}
\begin{document}
\maketitle

\ifdefined\hpcacameraready 
  \thispagestyle{plain}
  \pagestyle{plain}
\else
  \thispagestyle{plain}
  \pagestyle{plain}
\fi

\newcommand{\hpcaheight}{0mm}
\ifdefined\eaopen
\renewcommand{\hpcaheight}{12mm}
\fi


\begin{abstract}
  Graphics rendering that builds on machine learning and radiance fields is gaining significant attention due to its outstanding quality and speed in generating photorealistic images from novel viewpoints. 
  However, prior work has primarily focused on evaluating its performance through software-based rendering on programmable shader cores, leaving its performance when exploiting fixed-function graphics units largely unexplored.

  In this paper, we investigate the performance implications of performing radiance field rendering on the hardware graphics pipeline.
  In doing so, we implement the state-of-the-art radiance field method, 3D Gaussian splatting, using graphics APIs and evaluate it across synthetic and real-world scenes on today's graphics hardware.
  Based on our analysis, we present \name{}, which seamlessly integrates two innovations into graphics hardware to streamline the hardware pipeline for volume rendering, such as radiance field methods.
  First, we introduce native hardware support for early termination by repurposing existing special-purpose hardware in modern GPUs.
  Second, we propose multi-granular tile binning with quad merging, which opportunistically blends fragments in shader cores before passing them to fixed-function blending units.
  Our evaluation shows that \name{} greatly improves rendering performance, achieving up to a 2.78$\times$ speedup over the conventional graphics pipeline with negligible hardware overhead.
\end{abstract}

\putsec{intro}{Introduction}

The advent of graphics techniques that combine machine learning and radiance
fields has sparked significant interest in a new direction of representing 3D
scenes via implicit neural fields (e.g., NeRF~\cite{mil:sri20}) or explicit
rendering primitives that are also differentiable (e.g., 3D
Gaussian~\cite{ker:kop23}).
Compared to traditional methods that use meshes and textures, these radiance
field-based techniques allow us to capture intricate details in 3D scenes and
produce far more realistic images from new and unseen viewpoints, leading to
their rapid adoption and plug-in development for popular graphics engines such
as Unity~\cite{unity} and Unreal~\cite{unreal}.
In particular, the graphics community is actively exploring the Gaussian
splatting~\cite{ker:kop23} technique due to its ability to generate
high-fidelity images with much faster rendering speeds than other radiance
field methods.

The core of the impressive rendering performance of Gaussian splatting lies in
the use of explicit \emph{rasterization} primitives. While Gaussian primitives
can, in principle, be rendered through the hardware graphics pipeline using
graphics APIs such as OpenGL~\cite{opengl}, Vulkan~\cite{vulkan}, and
Direct3D~\cite{d3d}, prior work has primarily focused on assessing their
performance based on software-based rendering~\cite{ker:kop23,rad:ste24} or
building a specialized accelerator~\cite{lee:lee24}, thereby leaving the
potential of exploiting graphics-specific hardware largely unexplored.

In this paper, instead of developing a dedicated accelerator, we explore
leveraging fixed-function hardware in GPUs to improve the rendering efficiency
of Gaussian splatting on commodity GPUs.
In doing so, we first implement Gaussian splatting rendering using an industry
standard graphics API and evaluate it on modern desktop and edge GPUs.
Our implementation shows that hardware-based radiance field rendering generally
offers better or comparable performance compared to state-of-the-art
software-based rendering, though this can vary slightly depending on the scenes
and hardware configurations, such as the number of programmable shader cores or
fixed-function raster operation (ROP) units in GPUs. 
However, we also observe that the existing hardware graphics pipeline falls
short of efficiently performing Gaussian splatting rendering.
This is because Gaussian splatting is a volume rendering technique that
produces a pixel color by accumulating a huge amount of \emph{transparent}
primitives (i.e., Gaussians) into the color buffer through per-pixel
blending.

\begin{figure}[t]
  \centering
  \includegraphics[clip,trim={0in 0.25in 0.05in 0in}, width=\columnwidth]{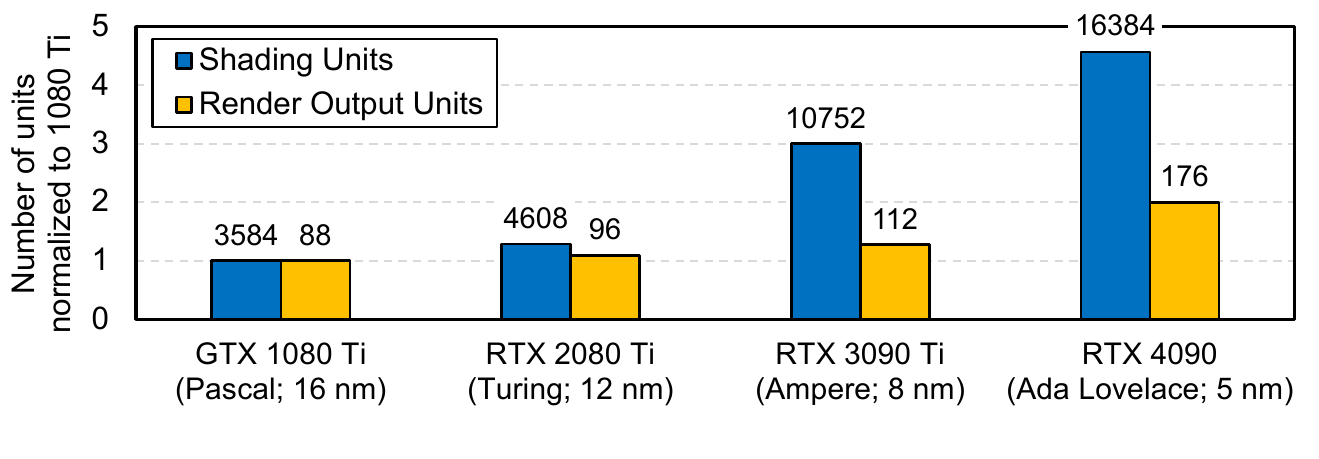}
  \caption{Number of shader cores and render output units in the recent generations of top-of-the-line NVIDIA desktop GPUs. 
  The labels on the bars indicate the absolute numbers for each.
  }
  \vspace{-0.20in}
  \label{fig:shader-vs-rop}
\end{figure}

In contrast, today's graphics hardware is primarily designed for mesh-based
rendering with mostly \emph{opaque} geometry. Although GPUs have seen a gradual
increase in the number of ROP units over successive generations, the growth has
been relatively modest, as shown in~\figref{shader-vs-rop}. 
For conventional mesh-based rendering, which typically involves one or only a
few fragments per pixel, ROP units rarely become the pipeline bottleneck.
However, volume rendering puts substantially more pressure on ROPs due to the
blending of hundreds of fragments per pixel.
Moreover, a commonly used optimization technique for volume rendering,
\emph{early ray termination}, is not natively supported in graphics hardware,
further limiting the potential for maximizing rendering performance.

Unfortunately, software-based optimizations aimed at reducing the ROP pressure
or supporting early termination do not provide much performance benefit.
For example, to alleviate the ROP pressure, one could use an extension feature
available in certain graphics APIs (e.g., OpenGL's \texttt{\small
ARB\_fragment\_shader\_interlock}) that allows for atomic pixel blending within
fragment shaders. However, this \emph{in-shader} blending approach can lead to
a significant drop in rendering performance due to the overhead of lock
acquisition~(\S\ref{ssec:in-shader-blending}).
Also, previous research has explored multi-pass rendering~\cite{kru:wes03} to
perform early termination using graphics APIs, but this approach offers limited
speedups or can even degrade performance due to the overhead of invoking
multiple intermediate rendering passes~(\S\ref{ssec:multipass}).

To this end, we present \name{} (Volume Rendering Pipeline), which features two
innovations that streamline the hardware graphics pipeline to better support
volume rendering workloads such as radiance field rendering.
The first is native hardware support for early termination by leveraging
existing special-purpose units (i.e., stencil test hardware) in contemporary
GPUs with minimal extension.
Our key observation is that both stencil test and early termination share
a similar purpose, so we can \emph{repurpose} the stencil test hardware for
checking early termination with negligible changes to ROPs.
The second is multi-granular tile binning with quad merging, which reduces the
number of blending operations performed in ROPs.
A key insight is that we can perform \emph{opportunistic} blending of
multiple fragments within a warp before passing them to ROPs by leveraging the
associative property of the blending equation and changing the computation
order.

We implement \name{} on the Emerald simulator~\cite{gub:aam19}, which builds on
gem5~\cite{bin:bec11} and GPGPU-Sim~\cite{bak:yua09}, while making extensive
modifications to the baseline implementation to better model contemporary
NVIDIA-like GPUs based on our analysis on real graphics hardware.
Our evaluation shows that \name{} improves Gaussian rendering performance by
{2.07$\times$} on average compared to the baseline graphics pipeline.
In summary, we make the following contributions:
\begin{itemize}
\item To our knowledge, this is the \emph{first} work to identify the
challenges and inefficiencies associated with Gaussian-based radiance field
rendering on the hardware graphics pipeline using graphics APIs.

\item We present \name{}, a hardware graphics pipeline that features native
support for early termination and multi-granular tile binning with quad
merging, which substantially improves the performance of volume rendering
workloads such as radiance field rendering.

\item We implement OpenGL-based microbenchmarks and provide an analysis
on several key fixed-function units in modern graphics hardware.
\end{itemize}

\putsec{back}{Background}

This section first provides the background on 3D graphics rendering. It then
describes the state-of-the-art radiance field rendering method: 3D Gaussian
splatting.

\begin{figure}[t]
  \centering
  \includegraphics[width=0.95\columnwidth]{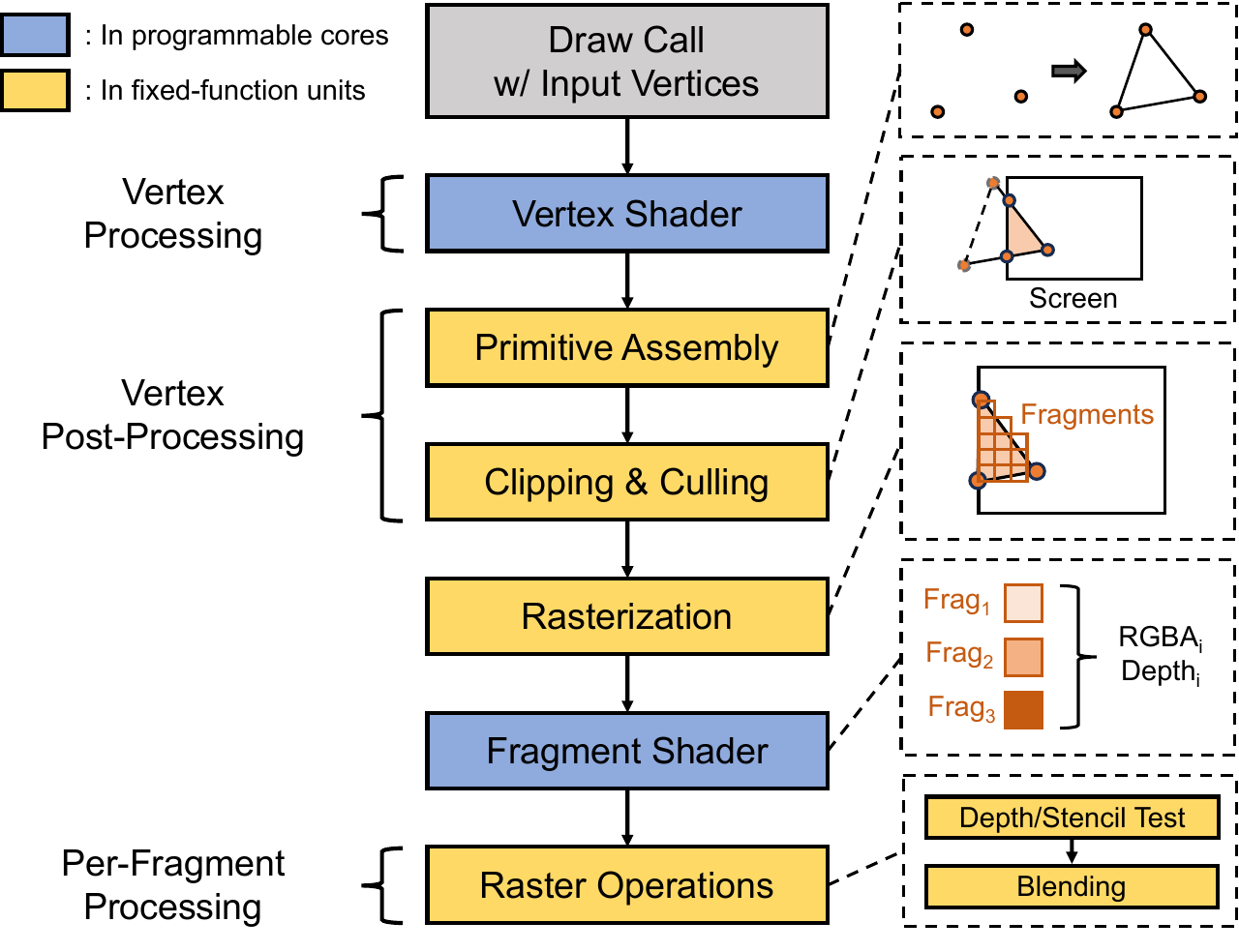}
  \caption{OpenGL rendering pipeline.}
  \vspace{-0.20in}
  \label{fig:opengl-pipeline}
\end{figure}

\putssec{gr}{Preliminaries on 3D Graphics Rendering}

\noindent \textbf{Graphics Pipeline.}
A 3D scene is rendered into a 2D image through a series of stages, which is
referred to as a \emph{graphics pipeline} or a \emph{rendering pipeline}.
To run the graphics pipeline, graphics software conventionally builds on
standard graphics APIs such as OpenGL~\cite{opengl}, Direct3D~\cite{d3d},
Vulkan~\cite{vulkan}, and Metal~\cite{metal}. Each API defines a set of
functions that process the operations in the rendering pipeline on graphics
hardware. 

\figref{opengl-pipeline} illustrates a high-level overview of the OpenGL
rendering pipeline. Other graphics APIs also employ a similar pipeline model.
The OpenGL pipeline can be largely divided into five stages: vertex shading,
vertex post-processing, rasterization, fragment shading, and per-fragment
processing. 
In hardware-based graphics rendering, each pipeline stage maps to either
programmable shader cores or fixed-function units in GPUs. 
Note that in software-based rendering, the operations in each stage are
executed entirely on the shader cores without using fixed-function hardware.

When a draw call is invoked with input vertices, the vertex shader\footnote{A
shader is a small program that runs on the shader cores.} transforms the
position of each vertex from 3D world space into clip space coordinates, which
will be further transformed into 2D screen positions and depth by
fixed-function hardware.
In the vertex post-processing stage, the vertices are then assembled into
primitives (e.g., triangles). In this stage, primitives outside the visible
space are excluded through a process known as \emph{view frustum culling},
and only the visible part of a primitive remains if part of the primitive is
outside the screen space.

The visible primitives are fed into a hardware rasterizer to identify the
pixels that overlap with them. The rasterizer produces \emph{fragments} for
each primitive; if a pixel is covered by multiple primitives, there will be
more than one fragment for the pixel.
Also, vertex attributes computed by the vertex shader are interpolated for each
fragment in this stage. 

Using the per-fragment data (e.g., pixel position, interpolated features) and
shared data (e.g., textures), the fragment shader computes and outputs a color
and an opacity (i.e., an RGBA value) for each fragment. 
In the final per-fragment processing stage, raster operations perform depth and
stencil tests. For the fragments that pass the tests, their RGBA colors are
blended or stored into the color buffer to generate the final pixel color.
It should be noted that the rendering pipeline can be implemented in
hardware with various optimizations, as long as the final pixel colors are
correctly produced.

\begin{figure}[t]
  \centering
  \includegraphics[width=0.95\columnwidth]{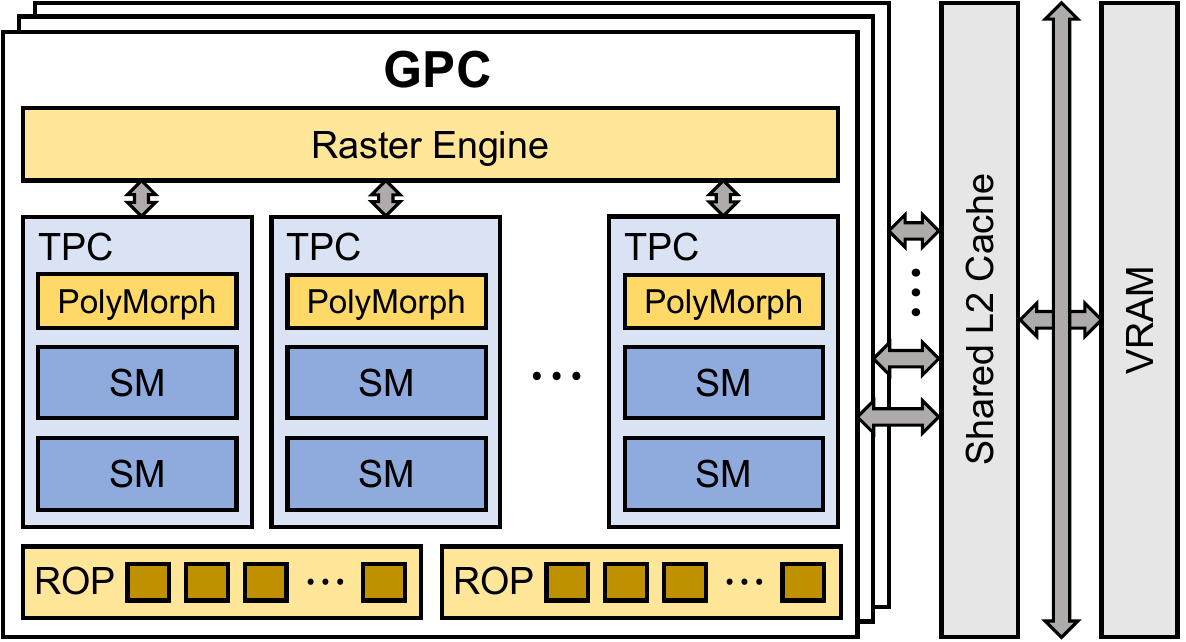}
  \caption{NVIDIA Ampere GPU architecture~\cite{ampere}.}
  \vspace{-0.20in}
  \label{fig:nvidia-gpu-arch}
\end{figure}

\myparagraph{Graphics-Specific Hardware in GPUs.}
As previously discussed, modern GPUs employ programmable shader cores that
execute different types of shader programs. Today, the shader cores are not
only accessible from graphics software but are also exposed to run
general-purpose programs through software frameworks such as CUDA and OpenCL.
Still, GPUs also feature graphics-specific hardware that facilitates the
execution of certain parts of the graphics pipeline, which is \emph{not}
accessible via general-purpose computing frameworks.

As shown in~\figref{nvidia-gpu-arch}, for example, an NVIDIA GPU includes
several special-purpose graphics units in addition to the programmable shaders
(i.e., Streaming Multiprocessor; SM).
Each Graphics Processing Cluster (GPC) includes a number of Texture Processing
Clusters (TPCs), each of which contains a PolyMorph Engine. The PolyMorph
Engine performs operations such as vertex fetching and viewport transformation,
and forwards the results to the Raster Engine~\cite{wit:kil11}.

The Raster Engine (rasterizer) sets up triangle edges using input vertex
positions and computes the pixel coverage of each triangle, a process called
rasterization. 
The fragments produced by the rasterizer are sent to the depth ($z$) test unit
(ZROP) if an early $z$-test is enabled.
This unit compares the depth of each fragment with the value in the $z$-buffer at
the same pixel position, discarding fragments that would ultimately fail the
late $z$-test conducted after fragment shading. By doing so, it prevents
unnecessary fragment shading computations.
After fragment shading in the shader cores (SM), the render output units
(ROPs), also known as raster operation units, perform blending or storing
operations while ensuring the proper ordering of fragments for the same pixel
location.

\myparagraph{Tile-Based Rendering.}
Most contemporary GPUs, including NVIDIA RTX, AMD Radeon, Intel Gen, and ARM
Mali, now use some variant of tile-based rendering (TBR).
When rendering an image using the hardware graphics pipeline, the screen space
is divided into a grid of screen tiles, each containing a block of pixels.
These tiles are assigned to the shader cores in the form of warps or thread
blocks.
For instance, NVIDIA GPUs split the screen space into a grid of
16$\times$16-pixel tiles, each of which is assigned to a specific GPC. 
This improves cache locality and reduces off-chip memory access during
rendering.
To achieve this, GPUs perform tile binning in hardware~\cite{gen11,lin:mor09}.
The fragments produced by the hardware rasterizer are grouped into bins based
on their tile IDs. These bins are then flushed to the shader cores when certain
conditions are met (e.g., a bin is full, a timeout occurs, or there is a lack
of available bins for new fragments with different tile IDs).
\secref{analysis} provides further analysis and discussion of fixed-function
units and tile-based rendering, based on microbenchmarking of modern GPUs.

\putssec{3dgs}{Radiance Field Rendering with Gaussian Splatting}

3D Gaussian splatting~\cite{ker:kop23} introduces a novel method that achieves
state-of-the-art rendering performance and quality by \emph{explicitly}
representing a scene with a set of anisotropic 3D Gaussians.
Each Gaussian is characterized by geometric properties, such as a position
(mean) coordinate $\mu$ and a 3$\times$3 covariance matrix $\Sigma$, as well as
visual properties, such as opacity $o$ and spherical harmonic (SH) coefficients
$sh$, to represent the view-dependent color of the Gaussian.

For training, given a sparse set of 2D images, an initial set of 3D points is
generated using a Structure-from-Motion (SfM) technique. These points serve as
the centers for the initial isotropic Gaussians.
During the training phase, the features of the Gaussians are updated
continuously based on their computed gradients. To better represent the fine
geometric details of the scene, the number of Gaussians increases as they are
cloned and split into smaller ones.

While a 3D Gaussian is mathematically defined as a continuous function
over the entire 3D space, Gaussian splatting models each Gaussian as an
ellipsoid for practical purposes.
During rendering, these 3D Gaussians are projected onto the 2D image plane as
ellipses, referred to as \emph{2D splats}. 
The splats are sorted by depth, from nearest to farthest relative to the given
viewpoint. The final pixel color ($\mathbf{{C}}$) is then computed using
$\alpha$-blending (\eqnref{volumerender}), which combines the colors
($\mathrm{\mathbf{c}_i}$) of overlapping splats in front-to-back order:
\begin{equation}
\small
\begin{aligned}
  \mathbf{{C}} = \mathrm{\sum\limits_{i=1}^{N}} &\mathrm{\alpha_i\mathbf{c}_i} \mathrm{\prod_{j=1}^{i-1}} \mathrm{(1-\alpha_j)}, \\
  \textrm{with}~~\mathrm{\alpha_i} = o_\mathrm{i} \cdot \mathrm{exp}(-\frac{1}{2}&({p'}-\mu')^T\Sigma'^{-1}({p'}-\mu')),
  \label{eqn:volumerender} 
\end{aligned}
\end{equation}
where ${p'}$ denotes the pixel position, and $\mu'$ and $\Sigma'$ represent the
mean and the covariance matrix of the 2D splat, respectively.

\putsec{motiv}{Motivation}

\putssec{}{3D Gaussian Splatting on Graphics Hardware}

To understand the performance implications of exploiting the hardware pipeline
for Gaussian splatting, we implement its rendering process using {OpenGL} and
evaluate it across synthetic and real-world scenes on mobile and desktop GPUs.

\begin{figure}[t]
  \centering
  \includegraphics[trim=0in 0.15in 0in 0in, clip=True, width=0.95\columnwidth]{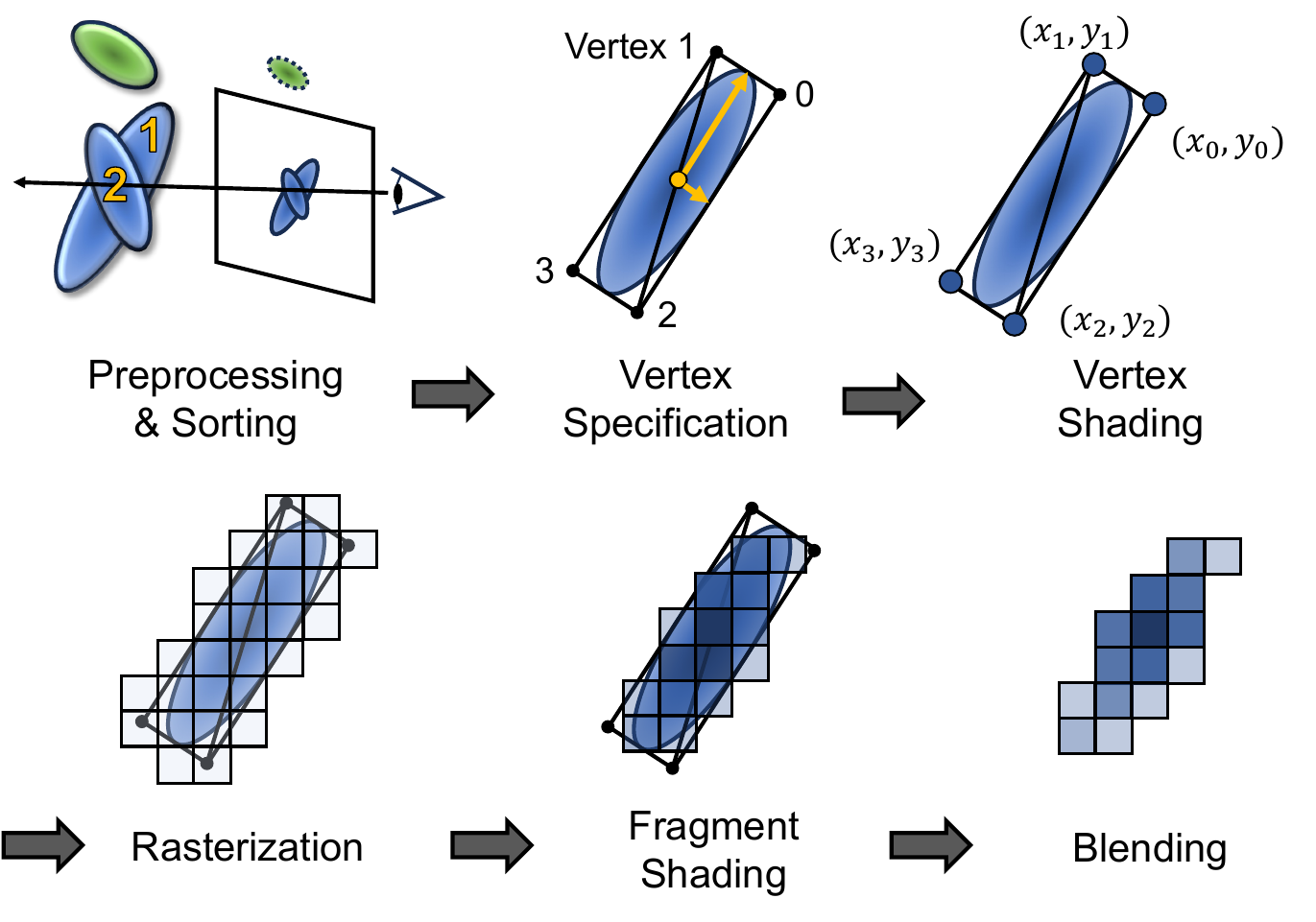}
  \caption{Gaussian splatting rendering via graphics APIs.}
  \vspace{-0.20in}
  \label{fig:opengl-3dgs}
\end{figure}

\myparagraph{Rendering 3D Gaussians via Graphics APIs.}
\figref{opengl-3dgs} illustrates how 3D Gaussians are rendered using graphics
APIs. To begin with, we implement custom CUDA kernels and use the NVIDIA CUB
library for the preprocessing and sorting steps, similar to the software-based
rendering~\cite{ker:kop23}. For the subsequent steps, we use {OpenGL} to
implement vertex and fragment shaders that execute on graphics hardware.

In the preprocessing and sorting steps, we first perform frustum culling to
exclude invisible Gaussians and calculate a depth value, which is the $z$-value
of the center of each Gaussian in camera space. We then project (i.e.,
``splat'') the Gaussians onto screen space and compute an RGB color of each
splat using SH coefficients and the viewing direction.
The splats are subsequently sorted by the depth values to perform alpha
blending front-to-back in a draw call.

Once we obtain a sorting order of the splats, we exploit fixed-function units
in graphics hardware for rendering. This requires each splat to be represented
as conventional graphics primitives (e.g., triangles). To achieve this, we
specify an Oriented Bounding Box (OBB)~\cite{got:lin96} that surrounds each
splat using two triangles with four vertices.
By using the center of the splat, two semi-axis vectors of the ellipse, and the
vertex indices, the vertex shader computes a 2D coordinate for each vertex in
screen space. Each vertex is also assigned the color and opacity values of the
corresponding Gaussian, which were previously computed during the preprocessing
step. 

Subsequently, the hardware rasterizer produces fragments that overlap with the
triangles. Afterward, the fragment shader computes the alpha value of each
fragment by evaluating a Gaussian function at the pixel position
(\eqnref{volumerender}). The fragments whose alpha values are small enough
(i.e., $\alpha<\frac{1}{255}$) are excluded from blending, which we call
\emph{alpha pruning} in this paper, and the remaining fragments are blended
into the corresponding pixels in the ROP units.

\begin{figure}[t]
  \centering
  \includegraphics[trim=0in 0.00in 0in 0in, clip=True, width=\columnwidth]{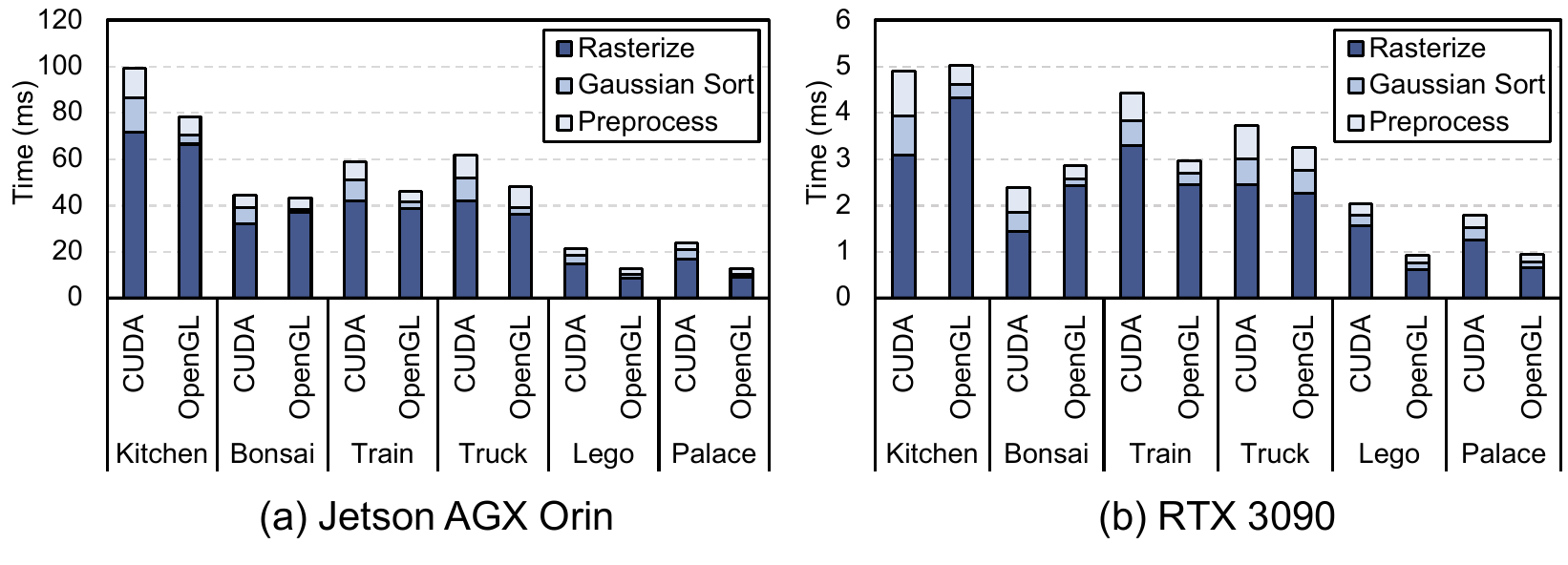}
  \caption{Performance comparison between software-based (CUDA) and hardware-based (OpenGL) graphics rendering on Jetson AGX Orin and RTX 3090. See~\tabref{workloads} for the details of each scene.}
  \vspace{-0.20in}
  \label{fig:cuda-vs-opengl}
\end{figure}

\myparagraph{Performance Analysis.}
\figref{cuda-vs-opengl} compares the performance of the state-of-the-art
software-based rendering using custom CUDA kernels~\cite{ker:kop23} and our
OpenGL-based rendering that exploits fixed-function graphics units.
To reduce the amount of work in rasterization, we use a tight OBB\footnote{The
Gaussian's boundary is defined where the alpha value is equal to
$\frac{1}{255}$.} that accounts for the opacity of each Gaussian in both CUDA
and OpenGL implementations. Specifically, for CUDA, this approach significantly
reduces the number of ineffective Gaussian-tile assignments, resulting in a
notable speedup in sorting and rasterization compared to the axis-aligned
bounding box (AABB)-based method used in the original CUDA implementation, as
discussed in previous studies~\cite{lee:lee24,rad:ste24}.
Note that this optimization does not alter the rendered image, as it only
reduces ineffective computations.

Results show that using fixed-function units generally offers better
performance than software-based rendering.
In software-based rendering, preprocessing is inefficient, requiring per-tile
buffers and duplicating depth and index data for Gaussians spanning multiple
tiles, which is time-consuming.
In contrast, hardware-based rendering eliminates the inefficiencies, as the
graphics hardware \emph{automatically manages} tiling and duplication.
This also leads to a reduction in sorting time, as we only need to sort the
entire Gaussians by depth values without the need of duplication and per-tile
sorting.
Furthermore, as noted in prior work~\cite{lee:lee24}, the lockstep execution of
threads in CUDA leads to \emph{ineffective} alpha computation during
rasterization.
Hardware-based rendering, operating at a finer granularity (e.g., a
2$\times$2-fragment quad) than a warp (i.e., 32 threads), allows for more
effective utilization of shader cores and ROPs, thereby improving performance
during the rasterization step.

While rendering using graphics hardware in GPUs is generally faster than
software-based rendering, it still falls short of real-time rendering,
particularly on mobile/edge devices with limited power and resources. 
As shown in~\figref{cuda-vs-opengl}(a), hardware-based rendering achieves less
than 25 FPS for real-world scenes.
In the following section, we further discuss our observations and the
inefficiencies of Gaussian-based rendering on the hardware pipeline.

\putssec{opportunities}{Observations and Opportunities}

\begin{figure}[t]
  \centering
  \includegraphics[trim=0in 0.00in 0in 0in, clip=True, width=\columnwidth]{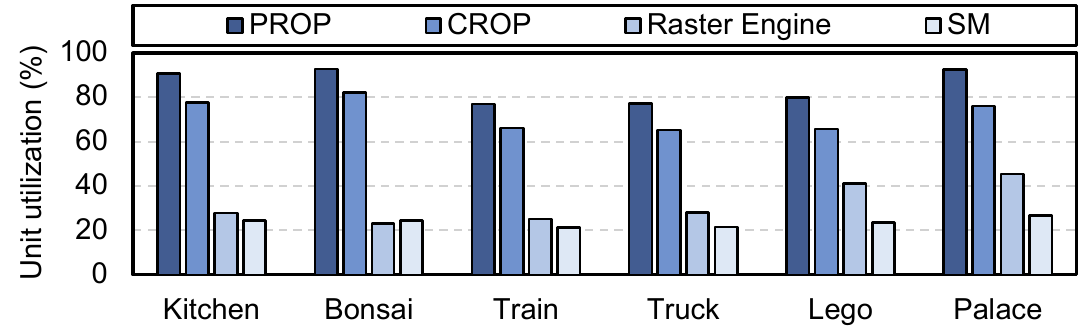}
  \caption{
    Throughput utilization of each hardware unit for OpenGL-based rendering, which is computed by $(\frac{\textrm{Measured
    Throughput}}{\textrm{Max Throughput}}){\times}100$. 
  }
  \vspace{-0.20in}
  \label{fig:unit-utilization}
\end{figure}

\textbf{\textit{Observation I: ROP Pressure and Inefficient Use of Shader
Cores.}}
\figref{unit-utilization} shows the average throughput utilization of several
key graphics hardware units during a draw call.
We observe that the rendering performance of Gaussian splatting is dictated by
the ROP units (i.e., PROP and CROP) rather than by the shader units (SM).
This is because there are a huge number of fragments to blend per pixel in
Gaussian splatting.
The PROP (Pre-ROP) orchestrates the flow of depth and color fragments for a
final pixel, while the CROP (Color ROP) performs blending in the order of
fragments received from the PROP.

In addition, the shader units are relatively underutilized because of two reasons.
First, the ROP units are the pipeline bottleneck, so the shader units are often
not fully utilized due to back pressure. Second, the vertex and fragment
shaders used for Gaussian splatting are relatively simple compared to
CUDA-based renderers. In detail, the vertex shader only computes the 2D screen
coordinate using the center of the splat and axis vectors, and the vertex
colors are shared for all vertices of the splat, which are already computed in
the preprocessing step. 
The fragment shader computes the alpha value of each fragment by applying a dot
product to a normalized pixel coordinate and performing an exponential
operation of the Gaussian function.
Compared to other shading programs that require lighting calculations and
texture fetching for each fragment~\cite{kru:wes03}, the shaders for Gaussian
splatting are computationally cheaper.

\begin{figure}[t]
  \centering
  \includegraphics[trim=0in 0.00in 0in 0in, clip=True, width=\columnwidth]{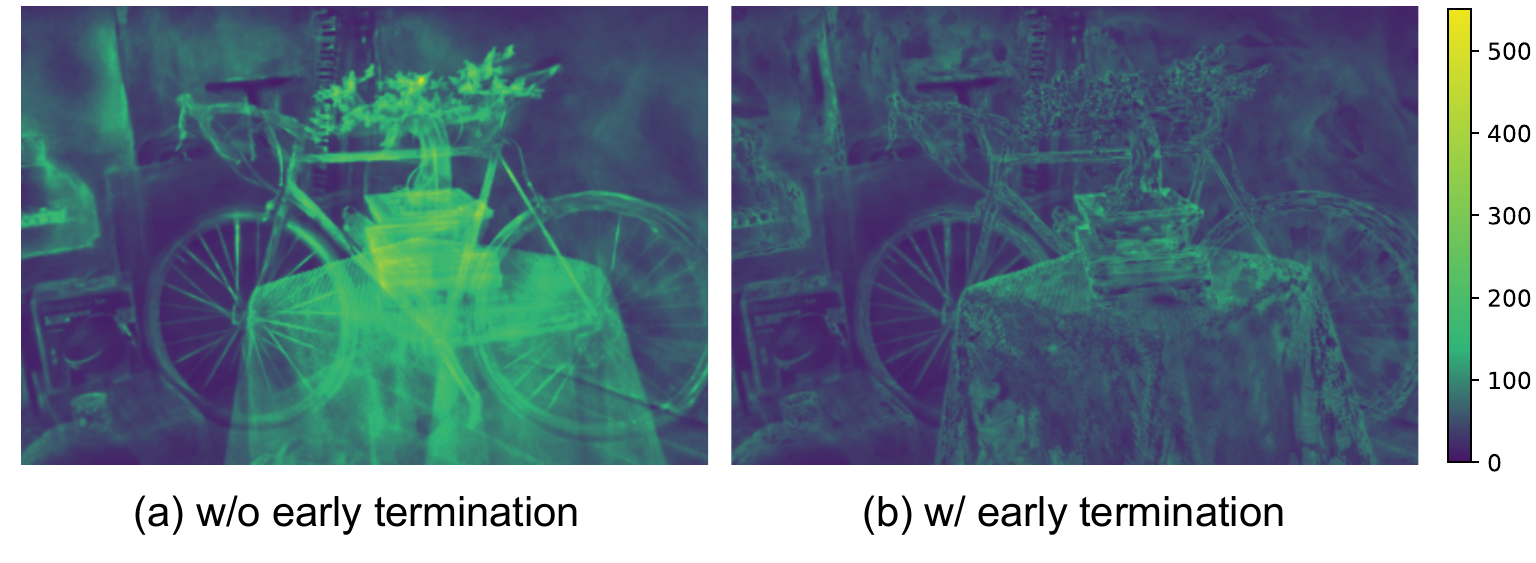}
  \caption{Number of fragments per pixel with and without early termination (scene: Bonsai).}
  \vspace{-0.20in}
  \label{fig:frag-heatmap}
\end{figure}

\myparagraph{\textit{Observation II: Challenges of Supporting Early
Termination.}}
Gaussian splatting is a volume rendering technique that accumulates RGBA colors
of fragments at the same pixel location \emph{front-to-back} to produce the
final pixel color.
Consequently, similar to other volume rendering methods, it can benefit from
\emph{early termination}, a widely-used optimization technique in the volume
rendering process.
\figref{frag-heatmap} illustrates the impact of early termination on Gaussian
splatting by comparing the number of fragments per pixel blended with
\emph{and} without early termination.
With early termination, if the alpha value of a pixel surpasses a predefined
threshold after blending, the subsequent fragments are discarded as they do not
make a noticeable contribution to the final pixel color.
As a result, this technique effectively reduces the amount of computation by
avoiding unnecessary shading and blending operations.

\begin{figure}[t]
  \centering
  \includegraphics[trim=0in 0.00in 0in 0in, clip=True, width=\columnwidth]{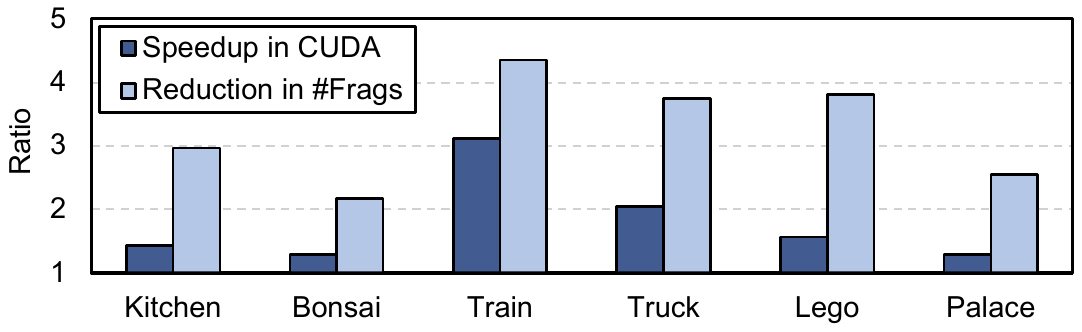}
  \caption{Speedup of CUDA-based rendering and the reduction in the number of fragments with early termination.}
  \vspace{-0.10in}
  \label{fig:ert-in-cuda}
\end{figure}

\begin{figure}[t]
  \centering
  \includegraphics[trim=0in 0.00in 0in 0in, clip=True, width=\columnwidth]{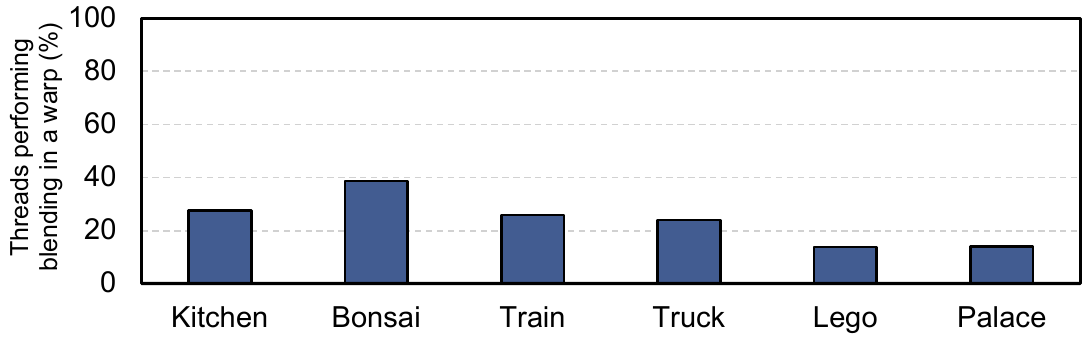}
  \caption{Average percentage of threads in a warp that participate in blending in software-based (CUDA) rendering.}
  \vspace{-0.15in}
  \label{fig:warp-divergence}
\end{figure}

\figref{ert-in-cuda} shows the speedup of CUDA-based rendering and the
reduction in per-pixel fragment count when using early termination.
In software-based rendering, while early termination helps improve rendering
performance, its full benefits are difficult to realize due to the lockstep
execution of threads working on different pixels.
In the worst case, even if only one thread (pixel) in a warp is not terminated,
all other threads in the warp still ineffectively consume shader cores
until the active thread finishes.
\figref{warp-divergence} shows the percentage of threads in a warp performing
blending operations for the evaluated scenes.
With the combined effects of alpha pruning and early termination, less than
40\% of threads perform effective work (i.e., blending) in a warp across all
scenes.

On the other hand, hardware-based rendering currently does \emph{not} natively
support early termination. However, adding this capability to graphics hardware
would significantly improve rendering performance for two reasons while better
exploiting its benefits compared to software-based rendering.
First, the hardware graphics pipeline performs fragment blending at a finer
granularity in ROPs (i.e., a quad; 2$\times$2 fragments) than a warp, thereby
enabling more effective utilization of compute units when early termination is
applied than software-based rendering.
Second, since volume rendering places substantial pressure on ROPs, reducing
the number of fragments sent to them would greatly streamline the hardware
pipeline by alleviating their processing burden.
To harness this benefit in the hardware pipeline, \ssecref{early-term-hw}
discusses our proposed architecture, which minimally extends the graphics
hardware by \emph{repurposing} fixed-function units, such as stencil test
hardware.

In the following section, we discuss the performance of in-shader blending and
software-based early termination via graphics APIs to demonstrate the need for
architectural support for volume rendering.

\putsec{sw-optim}{Software Optimizations and Limitations}

\begin{figure}[t]
  \centering     
  \includegraphics[width=\columnwidth]{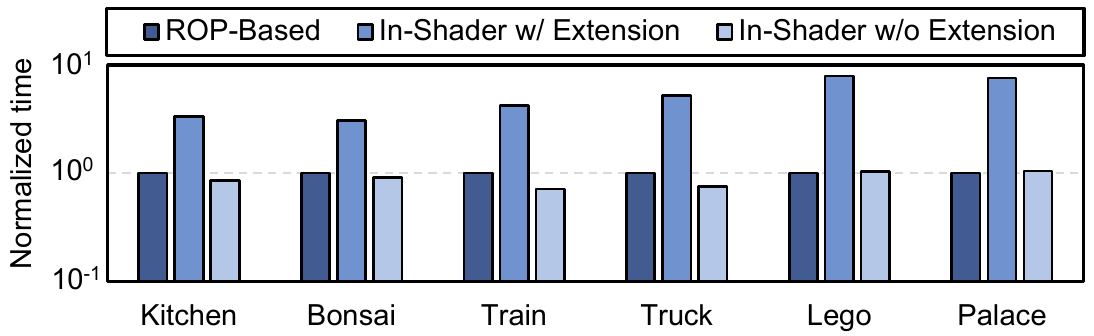}
  \caption {Normalized rasterization time of the ROP-based blending and
  in-shader blending with an OpenGL extension (log scale).}
  \vspace{-0.10in}
  \label{fig:in-shader-perf}
\end{figure}

\putssec{in-shader-blending}{Performing In-Shader Blending}

One approach to reduce the ROP pressure might be performing raster operations,
particularly pixel blending, in the fragment shader. While one may think that
this could take advantage of the high parallelism of shader cores, we cannot
na\"ively perform blending in the fragment shader because we need to ensure the
blending order of fragments to obtain accurate pixel colors. 
Even though graphics hardware forms thread blocks and dispatches them to the
shader cores in order, the fragment threads corresponding to the same pixel can
run \emph{out-of-order} due to warp scheduling policy and memory systems (e.g.,
cache hits/misses, memory controller). As a result, performing in-shader
blending without additional mechanisms does not prevent fragment threads from
violating the blending order and atomicity, thereby producing an incorrect
pixel color.

Recently, several standard graphics APIs offer an extension that enables the
use of a critical section in the fragment shader, such as \texttt{\small
GL\_ARB\_fragment\_shader\_interlock} in OpenGL and \texttt{\small
VK\_EXT\_fragment\_shader\_interlock} in Vulkan, where an order of lock
acquisition can be configured to follow the order of fragments.
We implement in-shader blending by modifying the fragment shader with the
OpenGL extension. After discarding the fragments with alpha values below the
threshold (i.e., $\epsilon$=$\frac{1}{255}$) through alpha pruning, alive
threads (fragments) attempt to acquire a lock by calling \texttt{\small
beginInvocationInterlockARB}.\footnote{We configure the extension so that
entering the critical section follows the order of the fragments.}
Inside the critical section, the pixel (RGBA) load, blend, and store operations
are performed in an atomic manner. By releasing the lock after the store, the
next fragment thread in line enters the critical section and reads the valid
pixel color.

We observe that in-shader blending for Gaussian splatting performs
substantially worse than ROP-based blending, as shown in
\figref{in-shader-perf}. 
The significant drop in performance is mainly due to the overhead of the
locking mechanism rather than the raster operations. 
As we can see, in-shader blending \emph{without} interlocking, which allows
fragment threads to run in parallel to access and update the same pixel in
non-deterministic order, performs close to or faster than ROP-based rendering.
However, it does not guarantee the exact blending results.
In summary, we conclude that it is necessary to use ROPs for blending in terms
of performance and preciseness in volume rendering.

\begin{algorithm}[t]
\caption{Multi-Pass Rendering with Early Termination}
\label{alg:multipass}
\renewcommand{\algorithmicrequire}{\textbf{Input:}}
\renewcommand{\algorithmicensure}{\textbf{Output:}}
\begin{algorithmic}[1]
  \small
  \REQUIRE N; Number of passes, G; Gaussians, View; Viewpoint
  \ENSURE  Pixels; RGBA pixel colors

  \STATE Pixels $\gets$ 0; Stencils $\gets$ 0
  \STATE Splats, Depths $\gets$ Preprocess(G, View)
  \STATE SortedIdx $\gets$ SortGaussians(Depths)
  \STATE Batches $\gets$ SplitIntoNBatches(Splats, SortedIdx)

  \FOR{each $i$ from 1 to N}
  \STATE \textbf{// Draw call with the $i$-th batch of splats}
  \STATE Pixels $\gets$ DrawBatch(Batches[$i$], Pixels, Stencils)
  \IF{$i <$ N}
  \STATE \textbf{// Stencil buffer update for terminated pixels}
  \STATE Stencils $\gets$ EarlyTerminate(Pixels.A, Stencils)
  \ENDIF
  \ENDFOR
\end{algorithmic}
\end{algorithm}

\putssec{multipass}{Software-Based Early Termination}

To demonstrate the effectiveness of hardware-based early termination, we first
implement software-based (OpenGL) early termination via conventional multi-pass
rendering approaches~\cite{kru:wes03}.
The key concept of the multi-pass approach for early termination is to render
an image with \emph{multiple} draw calls. If a pixel is terminated on the
$i$-th draw call due to an early termination condition (i.e., $\alpha$ $\geq$
0.996), we skip processing the fragments of the terminated pixel from the
$(i+1)$-th draw call using a stencil test.\footnote{The stencil test is one of
the raster operations, which controls fragment processing by comparing the
per-pixel stencil value in a stencil buffer with the predefined reference
value.}
As the original algorithm~\cite{kru:wes03} was devised for texture-based volume
rendering, we begin by describing our algorithm for Gaussian splatting.

\algref{multipass} outlines the procedure for multi-pass rendering with early
termination. Initially, the colors and stencil values of pixels are set to 0. A
stencil value of zero indicates that the corresponding pixel has not yet been
terminated. 
For $\mathrm{N}$-pass rendering, the algorithm first divides the depth-ordered
splats equally into $\mathrm{N}$ batches after preprocessing and sorting. It
then iterates through multiple passes, with each pass consisting of two draw
calls.
In the first call, we \emph{partially} update the colors of the pixels that
pass the stencil test (i.e., stencil value == 0). In the second call, we update
the stencil values for the pixels that meet the termination condition in the
first call.

To update the stencil value, the second call renders a screen-sized rectangle
composed of two triangles using a different fragment shader and stencil test.
The rasterizer produces a fragment for every pixel, which is sent to the
fragment shader. Each fragment in the shader loads the alpha of the pixel and
is discarded if the pixel is not terminated yet (i.e., $\alpha$ $<$ 0.996).
After the shading, the stencil test is performed for the non-discarded
fragments, which represent the terminated pixel locations up to this pass, and
updates the stencil values to 1.
Consequently, the fragments of early-terminated pixels are not shaded or
blended in subsequent passes, reducing the ROP pressure.

\begin{figure}[t]
  \centering     
  \includegraphics[width=\columnwidth]{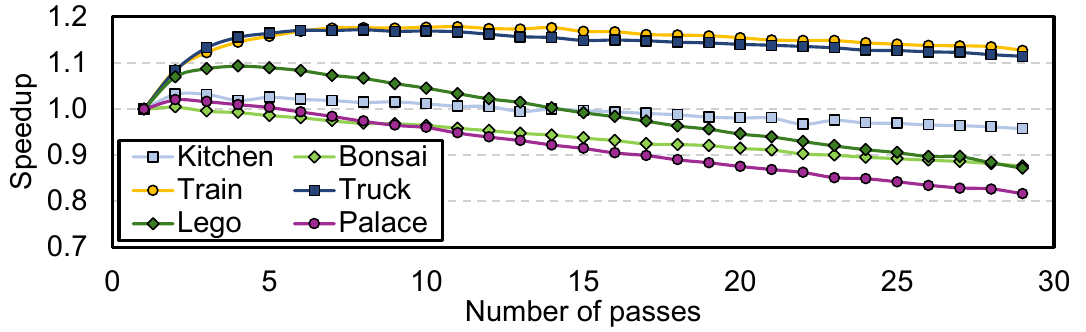}
  \caption {Performance of software-based early termination (OpenGL) with multi-pass
  rendering across the number of passes.} 
  \vspace{-0.20in}
  \label{fig:multipass-perf}
\end{figure}

\figref{multipass-perf} shows the speedup of software-based early termination
according to the number of passes. The baseline is the original OpenGL-based
rendering, where the number of passes ($\mathrm{N}$) is 1. The performance is
influenced by two factors: the reduction in the number of fragments due to
early termination and the overhead from additional draw calls for stencil
updates. While using a larger $\mathrm{N}$ can reduce fragments more by
checking early termination in a finer-grained manner, it also increases the
overall overhead of stencil updates.
As a result, for the scenes with low fragment reduction (e.g., Bonsai, Kitchen)
or already short rendering times (e.g., Lego, Palace), the performance
improvement is marginal, or it performs worse than the baseline due to the
overhead.
In large scenes with more fragment reduction (e.g., Train, Truck), we can
observe a speedup, but it does not reach its full potential due to the stencil
update overhead, which hardware-based early termination helps avoid.
In addition, the optimal number of passes varies depending on the scene and
viewpoint, making the software-based approach less practical. Support for
hardware-based early termination can also help address this issue.

\putsec{arch}{\name{}: Streamlining Graphics Hardware}

\putssec{overview}{Overview of Architecture and Execution Flow}

\begin{figure}[t]
  \centering
  \includegraphics[width=\columnwidth]{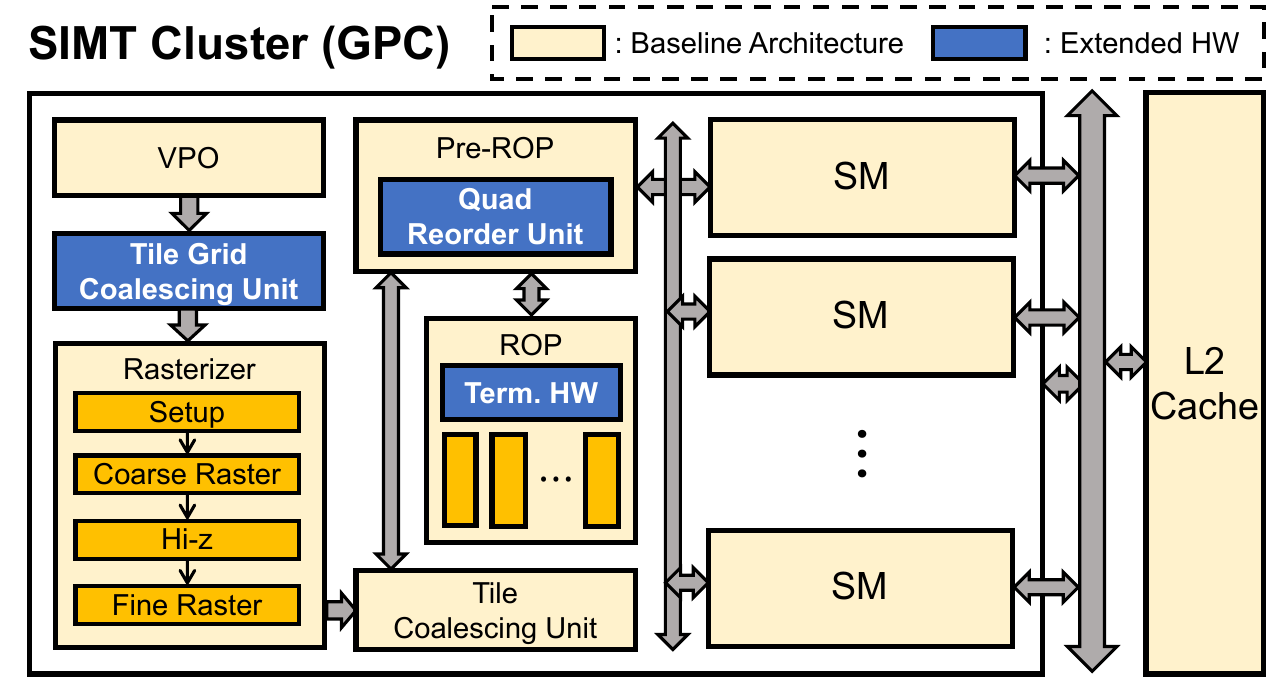}
  \caption{\name{} microarchitecture.}
  \vspace{-0.20in}
  \label{fig:vr-pipe-arch}
\end{figure}

\figref{vr-pipe-arch} provides an overview of the baseline GPU architecture
along with our hardware extensions for volume rendering. Because graphics
hardware is designed to support the standard graphics APIs, the high-level
microarchitecture is broadly similar across different hardware vendors, though
each may implement minor optimizations. Accordingly, we build our extensions on
a general graphics architecture commonly found in contemporary GPUs and
described in prior
literature~\cite{gub:aam19,tin:sax23,lin:mor09,rho:mol14,pur:mol13}, combined
with our analysis of real graphics hardware. Note that the hardware extensions
are generally applicable as they do not significantly alter the standard
hardware pipeline.

When input vertices are assigned to a SIMT cluster, the shader cores fetch
vertex data from memory, perform vertex shading, and output the resulting
vertices with attributes. 
Instead of passing the actual attribute data, a pointer to the attributes is
sent to the next pipeline stage after the attributes are stored in a designated
memory region in the L2 cache.\footnote{In NVIDIA terminology, this memory
region is called the Circular Buffer (CB), and the corresponding cache lines
are marked as no-evict~\cite{rho:mol14}.}

The vertices are then assembled as triangle primitives in the vertex processing
and operations (VPO) unit. For each primitive, the VPO unit also identifies
intersecting \emph{screen} tiles (e.g., a block of 16$\times$16 pixels) by
computing the bounding box of each primitive. Since each SIMT cluster is
exclusively responsible for a set of screen tiles, the primitives are then
distributed across SIMT clusters through a crossbar to run the subsequent
pipeline stages. If a primitive intersects multiple tiles, it is sent to
multiple different clusters and processed independently.

The tile grid coalescing (TGC) unit then manages these primitives at a coarser
granularity than a screen tile, which we call a \emph{tile grid} (e.g.,
4$\times$4 screen tiles).
Each bin in the TGC unit manages a tile grid and collects primitives that
intersect with any of the cluster's screen tiles within that grid.
When a bin has collected a sufficient number of primitives for the tile grid,
the TGC unit flushes the bin.

Once a TGC bin is flushed, the rasterizer performs rasterization for every
primitive in four sequential steps: setup, coarse raster, hierarchical-z (Hi-z)
test, and fine raster.
The setup unit first computes edge equations of the primitive using vertex
coordinates. The coarse raster then identifies intersecting \emph{raster} tiles
(e.g., 8$\times$8 pixels) within a screen tile.
The Hi-z test is performed on each raster tile, and only the raster tiles that
pass the test are sent to the next pipeline stage.
Finally, the fine raster unit checks if each pixel in the raster tile is
covered by the primitive. The pixels within the primitive are then assembled as
2$\times$2-fragment quads.

The tile coalescing (TC) unit manages a set of TC bins, each collecting the
quads for the \emph{same} screen tile.
It aggregates the quads from the fine raster unit into the corresponding bin
and flushes them to the Pre-ROP (PROP) hardware, which controls the blending
order between multiple quads. The flush occurs when one of three conditions is
met.
First, the bin is full.
Second, all bins are occupied and a quad from a new tile arrives (the oldest
bin is flushed).
Third, pre-determined cycles have elapsed after the last incoming quad.

When the bin is flushed, the quads are sent to the depth-stencil ROP (ZROP) for
an early termination check before fragment shading.
This hardware support for early termination operates at the \emph{fragment}
level during a \emph{single} draw call, thereby realizing most of its potential
compared to CUDA-based or OpenGL-based implementations.
The quads with at least one fragment that passes the early termination test are
sent back to the PROP. A quad reorder unit in the PROP then identifies the
overlapping quads that can be \emph{merged} in the shader cores and sets the
necessary flags. These quads are then dispatched to the shader cores as warps
for fragment shading, where quad merging is performed via warp shuffling.

After the shading, the surviving fragments from alpha pruning and quad merging
are sent to the color ROP (CROP) for pixel blending.
The termination hardware in the ROP checks if the termination condition is met
after blending and updates termination information as needed.
In the following subsections, we explain the architectural details of
hardware-based early termination and multi-granular tile binning with quad
merging.

\begin{figure}[t]
  \centering
  \includegraphics[clip,trim={0in 0in 0in 0in}, width=0.95\columnwidth]{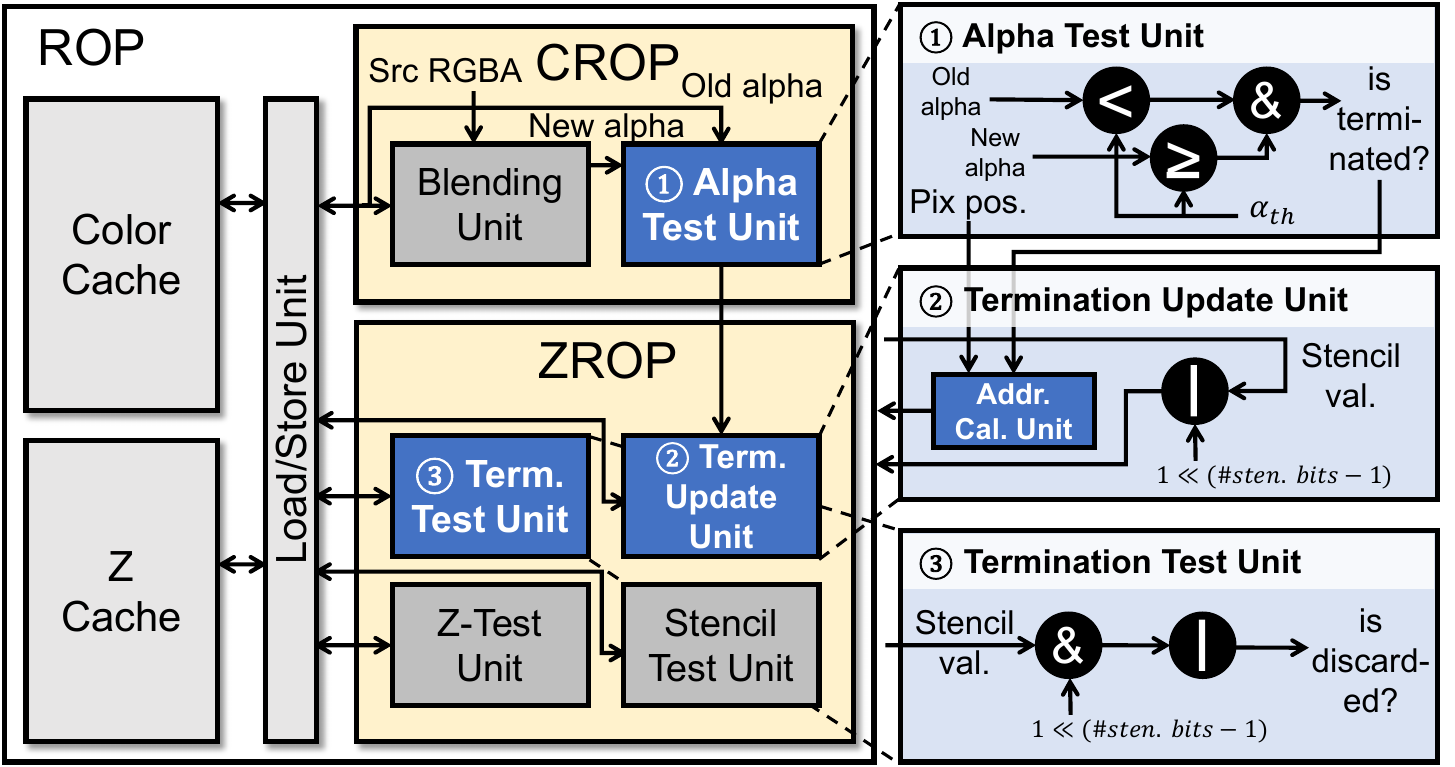}
  \caption{Early termination hardware in ROP.}
  \vspace{-0.20in}
  \label{fig:early-term-hw}
\end{figure}

\putssec{early-term-hw}{Hardware Support for Early Termination}

\figref{early-term-hw} shows hardware extensions to enable early termination
in the graphics pipeline. Early termination and stencil test share a similar
purpose: eliminating fragments that do not affect the final rendered output
before shading and blending. In this sense, the early termination unit can be
architected in the same stage where the stencil test unit is placed.

Our key insight is that multi-bit (e.g., 8-bit) stencil data per pixel is
already used for the stencil test, and we only need one bit data for checking
if the pixel is early terminated.
In most cases, only a few bits of the stencil value are used for the stencil
test by masking (e.g., \texttt{\small glStencilMask(0x01)}), and the remaining
bits can be used for the early termination test. 
By repurposing the most significant bit (MSB) of the stencil value for the
termination check, both early termination and the stencil test can be supported
harmonically. 

For example, when an 8-bit stencil value is used, the MSB serves as a
termination flag, while the remaining 7 bits are used for the stencil test.
Initially, the MSB is set to 0, allowing fragments of the pixel to pass the
termination test and blend with the pixel color.
Once the pixel's alpha value is sufficiently accumulated, the MSB is set to 1,
marking the pixel as terminated.
All subsequent fragments for the pixel are then discarded before shading.
As such, the early termination unit can be implemented with negligible overhead
by leveraging the existing stencil buffer. 

We add three lightweight computational units to the ROP. First, after blending
the pixel color (destination RGBA) with the shaded fragment color (source RGBA)
in CROP, the alpha test unit determines if the early termination condition is
met by checking if the alpha exceeds a predefined threshold by this fragment.
Instead of comparing only the accumulated alpha value, we also check if the
previous alpha does not exceed the threshold.
This is because using only the former condition would unnecessarily generate
more termination bit update requests to ZROP, which leads to bandwidth
contention to the z-cache and an increase in latency for the termination test.
If the pixel is newly terminated, the alpha test unit then sends a termination
signal to ZROP with a pixel coordinate.

In ZROP, the termination update unit is triggered by the termination signal and
updates a termination bit to 1. After the address calculation with the pixel
coordinate, it first loads the stencil value and sets the termination bit using
a bitwise OR operation. The updated stencil value of the terminated pixel is
then written back to the z-cache.

The early termination test is performed for the fragments that are flushed from
the TC unit. By conducting the test, we can effectively discard fragments of
the pixels that are already early-terminated and no longer need to be blended.
Consequently, this early termination test reduces the amount of work for shader
cores and ROPs, easing the burden of fragment shading and blending,
respectively.

\putssec{quad-merging}{Reducing ROP Workload by Quad Merging}

It is costly and challenging to increase ROP throughput by adding more blending
units to the SIMT cluster and increasing ROP cache bandwidth.
Instead, we leverage shader cores for blending, which offer high compute
throughput yet are underutilized in Gaussian rendering.
Our key insight is that volume rendering has an \emph{associative} property, so
we can alter the order of fragment blending---though not arbitrarily.
\begin{equation}
\small
\begin{aligned}
  \mathbf{{C}} = \mathrm{\sum\limits_{i=1}^{3}} \mathrm{\alpha_i\mathbf{c}_i} \mathrm{\prod_{j=1}^{i-1}} \mathrm{(1-\alpha_j)} 
  =& f_{fb} (f_{fb}(\mathbf{c}_{\mathrm{pm},1}, \mathbf{c}_{\mathrm{pm},2}), \mathbf{c}_{\mathrm{pm},3}),
  \label{eqn:a-blending-order}
\end{aligned}
\end{equation}
\begin{equation*}
\text{ where}
\footnotesize
  \text{$
  ~\mathbf{c}_{\mathrm{pm,i}}=\mathrm{(\alpha_i r_i,\alpha_i g_i}\mathrm{, \alpha_i b_i, \alpha_i)},
  ~f_{fb}(\mathbf{c}_{1}, \mathbf{c}_{2}) = \mathbf{c}_{1} + (1 - \alpha_1) \mathbf{c}_{2}.
  $
  }
\end{equation*}

\eqnref{a-blending-order} shows an example of blending two fragments
$\mathbf{c}_2$ and $\mathbf{c}_3$ into a (pre-multiplied alpha) pixel RGBA
color $\mathbf{c}_{\mathrm{pm},1}$. 
In the hardware pipeline, fragments are sent from PROP to CROP in front-to-back
order.
Since the front-to-back alpha blending equation $f_{fb}$ is associative, i.e.,
$f_{fb}(f_{fb}(\mathbf{c}_{1}, \mathbf{c}_{2}), \mathbf{c}_{3}) =
f_{fb}(\mathbf{c}_{1}, f_{fb}(\mathbf{c}_{2}, \mathbf{c}_{3}))$, we can
\emph{partially} change the computation order of volume rendering.
This allows for \emph{opportunistically} blending fragments in shader cores,
thereby reducing the number of blending operations by ROPs without
affecting the final pixel color; note that we cannot obtain the correct color
with $f_{fb}(\mathbf{c}_2, f_{fb}(\mathbf{c}_1, \mathbf{c}_3))$ though.

\begin{figure}[t]
  \centering
  \includegraphics[clip, trim={0cm 0.0cm 0cm 0cm}, width=0.95\columnwidth]{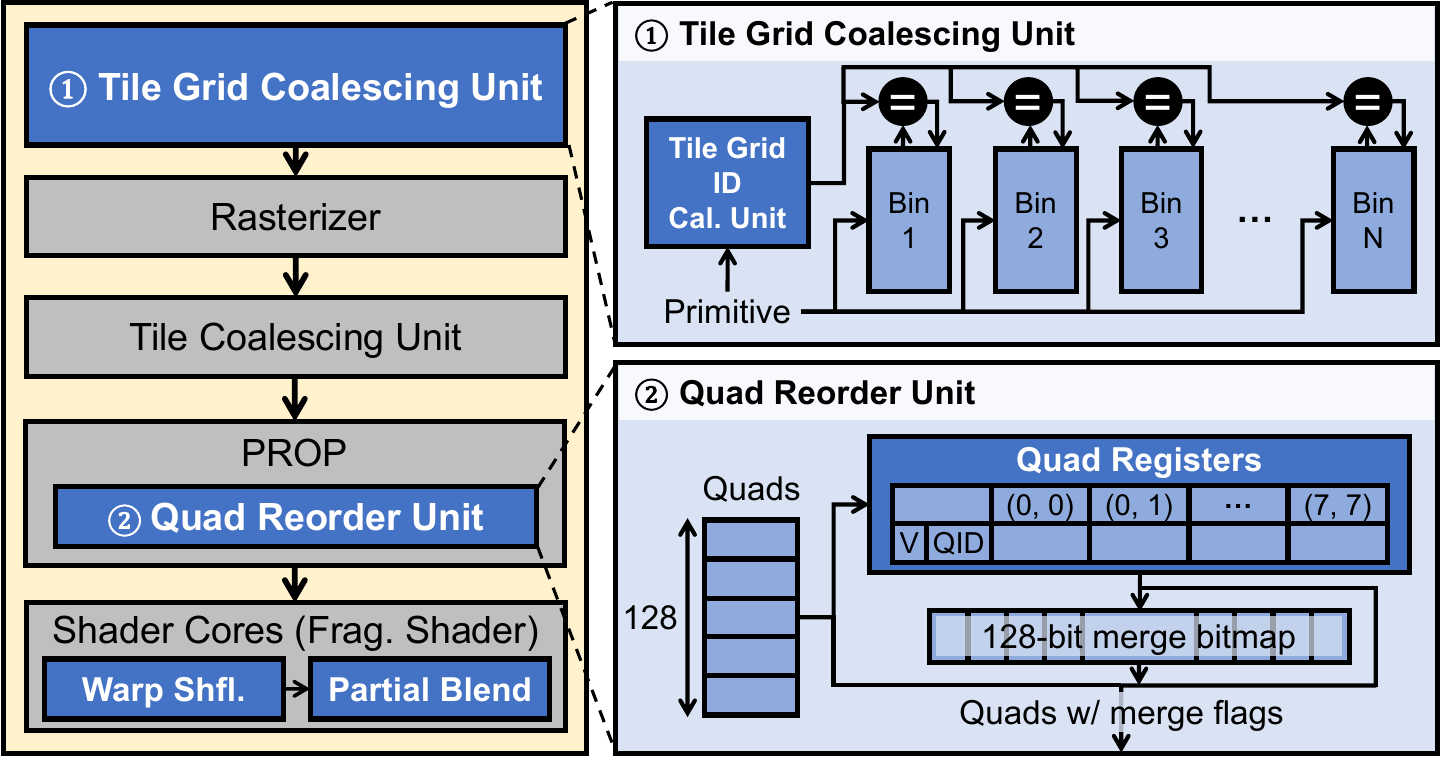}
  \vspace{-0.03in}
  \caption{Overall hardware pipeline for quad merging.}
  \vspace{-0.19in}
  \label{fig:quad-merging-hw}
\end{figure}

\figref{quad-merging-hw} shows the hardware pipeline for quad merging. Since
the conventional hardware graphics pipeline processes quads as the smallest
unit of granularity, our goal is to identify overlapping quads within a warp
and \textit{merge} them into a single quad by performing partial blending in
the shader cores. To do so, we propose hardware and software extensions
designed to merge the quads as much as possible.

On the hardware side, we introduce a \emph{tile grid} coalescing (TGC) unit and
a quad reorder unit.
Without the TGC unit, the TC unit simply gathers quads that belong to the same
screen tile into the TC bin, and the subsequent quad reorder unit attempts to
find overlapping quads.
However, the limited number of TC bins causes them to be frequently flushed
before collecting a sufficient number of quads to be merged, particularly when
primitives are large or spatially distributed.
To increase the opportunity to merge quads, the TGC unit first gathers the
primitives that intersect the same \emph{tile grid} into the TGC bin. 
Note that the \emph{tile grid} is larger than a screen tile but is still
smaller than the region that all TC bins in the TC unit can cover.
When the TGC bin is flushed, the rasterizer processes the primitives within the
\emph{tile grid} and generates quads that can be handled by a set of TC bins,
thereby mitigating the premature flushing of the TC bins.

When the TC bin is flushed, the quad reorder unit (QRU) identifies the
overlapping quads and reorders them to launch together in the same warp.
The input quads are stored in the buffer and assigned sequential quad IDs
(QIDs) from 0 to 127.
For overlap detection, the QRU maintains 64 8-bit (consisting of a valid bit
and 7-bit quad ID) registers, each corresponding to a relative quad location in
a screen tile from position $(0,0)$ to $(7,7)$.
The unit sequentially examines each quad starting from QID 0 and stores its QID
in the register corresponding to its location.
If the register already contains a valid QID, this indicates an overlap between
two quads.
Upon detecting overlapping quads, the QRU adds the pair into a warp with a
merge flag indicating that they need to be merged in the fragment shader.
The unit also manages a 128-bit bitmap where each bit represents whether the
quad will be merged.
After examining all quads, the QRU fills the warps with the quads that will not
be merged using the bitmap.

\begin{figure}[t]
  \centering
  \includegraphics[width=0.95\columnwidth]{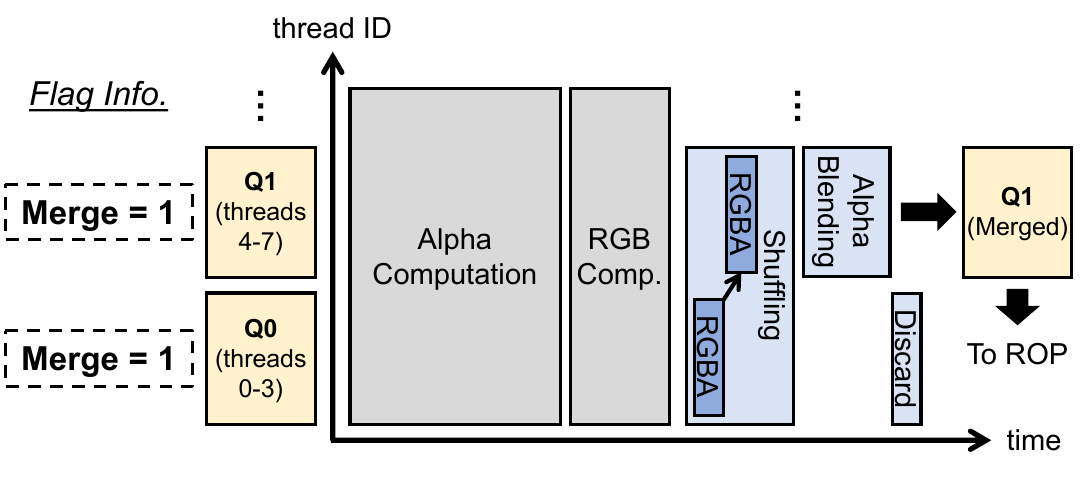}
  \caption{Execution timeline of fragment shader with quad merging.}
  \vspace{-0.15in}
  \label{fig:merging-flow}
\end{figure}

On the software side, we propose an extension that can be added to the original
shader for quad merging.
\figref{merging-flow} illustrates the execution flow of the fragment shader
with the proposed software extension.
After completing the original fragment shading operations, the threads marked
with a merge flag engage in partial blending. 
In the figure, for instance, two quads at quad offset 0 (threads 0-3) and 1
(threads 4-7) need to be merged.
Since two quads to be merged are always adjacent in the warp by the quad
reorder unit, the quad at $(2n+1)$ quad offset retrieves the fragment colors
from the quad at $2n$ offset using warp shuffling and blends them with its own.
Ultimately, a single merged quad is produced and sent to the ROP for final
blending into the color buffer.
By leveraging the high bandwidth of warp shuffling and the high compute
throughput of shader cores, we can increase effective blending throughput,
which would otherwise be limited by the number of ROP units.

\putsec{eval}{Evaluation}
\putssec{method}{Experimental Methodology}

\begin{table}[t]
  \centering
  \caption{Simulation configuration.}
  \label{tab:sys-config}
  \resizebox{0.70\columnwidth}{!}{%
    \begin{tabular}{|c|c|}
      \hline
      \multicolumn{2}{|c|}{\textbf{GPU}} \\
      \hline
      \# GPC & 1 \\
      \hline
      \# SIMT Cores & 16 (1024 CUDA Cores) \\
      \hline
      SIMT Core Freq. & 612 MHz \\
      \hline
      Lanes per SIMT Core & 64 (4 warp schedulers)\\
      \hline
      L1D/T & 48 KB, 128B line \\
      \hline
      Shared L2 & 4 MB, 128B line (sectored) \\
      \hline
      CROP Cache Size & 16 KB, 128B line (sectored) \\
      \hline
      Raster Tile Size & 8$\times$8 pixels \\
      \hline
      Tile Grid Size & 64$\times$64 pixels (4$\times$4 tiles) \\
      \hline
      \# of TGC Bins & 128 \\
      TGC Bin Size & 16 primitives \\
      \hline
      \# of TC Bins & 32 \\
      TC Bin Size & 128 quads \\
      \hline
      ROP Throughput & 2 quads/cycle (RGBA16F)\\
      \hline
      DRAM & LPDDR3-1600 (16-channel) \\
      \hline
    \end{tabular}
  }
\end{table}


\noindent\textbf{Simulation Infrastructure.}
To evaluate the rendering performance of graphics workloads, we use the Emerald
simulator~\cite{gub:aam19} that builds on the gem5 and GPGPU-Sim simulators. 
Emerald models an SoC system that comprises both CPU and GPU cores.
We use \emph{standalone} (GPU) mode and implement our hardware extensions on
the baseline GPU architecture that models the hardware stages in contemporary
GPUs.
Because Emerald does not implement the hardware units that we need to build on,
we make extensive modifications to the codebase, based on our analysis on
graphics hardware, which we discuss in~\secref{analysis}.
We set the GPU frequency and DRAM bandwidth to match those measured using
Jetson AGX Orin in 30W power mode.
We measure the execution cycles between the start of a draw call and the end of
raster operations, which captures the performance of an end-to-end graphics
pipeline.
\tabref{sys-config} shows the system configuration used in this work.


\begin{table}[t]
  \caption{Evaluated workloads.}
  \label{tab:workloads}
  \resizebox{\columnwidth}{!}{%
    \begin{tabular}{ccc}
      \toprule
      \multicolumn{1}{c}{\textbf{Dataset}} &
      \multicolumn{1}{c}{\textbf{Scene (Resolution / \#Gaussians)}} & \multicolumn{1}{c}{\textbf{Type}} \\
      \midrule
      \multirow{2}{*}{Mip-NeRF 360~\cite{bar:mil22}} &
      \multicolumn{1}{c}{Kitchen (1552$\times$1040 / 1.85M)} & \multirow{2}{*}{Real World \& Indoor}  \\
      \cmidrule(lr){2-2}
      & 
      \multicolumn{1}{c}{Bonsai (1552$\times$1040 / 1.24M)} &  \\ 
      \cmidrule(lr){1-3}
      \multirow{2}{*}{Tanks\&Temples~\cite{kna:par17}} & 
      \multicolumn{1}{c}{Train (980$\times$545 / 1.03M)} & \multirow{2}{*}{Real World \& Outdoor}  \\
      \cmidrule(lr){2-2}
      & 
      \multicolumn{1}{c}{Truck (979$\times$546 / 2.54M)} & {} \\
      \cmidrule(lr){1-3}
      \multirow{1}{*}{Synthetic-NeRF~\cite{mil:sri20}} &
      \multicolumn{1}{c}{Lego (800$\times$800 / 358K)} & \multirow{2}{*}{Synthetic}  \\
      \cmidrule(lr){1-2}
      \multirow{1}{*}{Synthetic-NSVF~\cite{liu:gu20}} &
      \multicolumn{1}{c}{Palace (800$\times$800 / 327K)} & {} \\
      \bottomrule
    \end{tabular}
  }
\end{table}


\myparagraph{Workloads.}
\tabref{workloads} shows the workloads that we use to evaluate \name{}.
We carefully select six widely-used scenes from published
datasets~\cite{bar:mil22,kna:par17,mil:sri20} to evaluate real-world and
synthetic scenes with varying complexity.
For each scene, we train the model for 30K iterations, which results in
approximately {300K to 2.5M} Gaussians per scene.
We generate the trace of OpenGL ES API calls using apitrace~\cite{apitrace} and
simulate the rendering at the full image resolution specified in the table.
For the evaluated viewpoints, up to 900K Gaussians are within the viewing
frustum.

\putssec{results}{Performance Results}

\begin{figure}[t]
  \centering
  \includegraphics[width=\columnwidth]{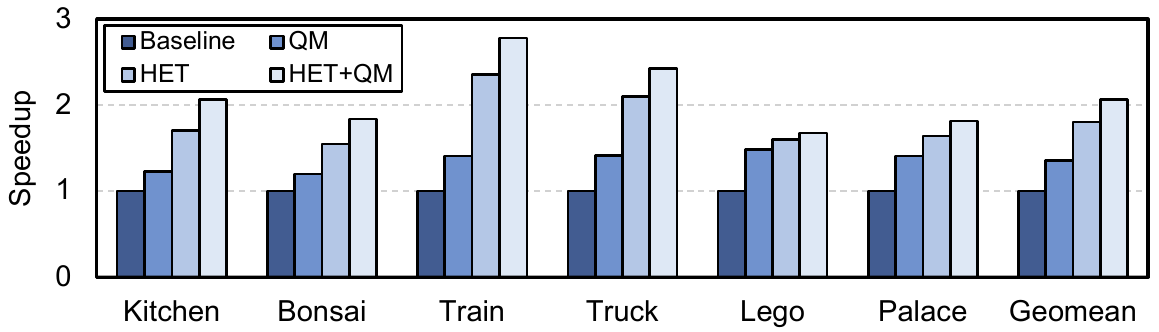}
  \caption{{Speedup of \name{} over the baseline GPU.}}
  \vspace{-0.20in}
  \label{fig:perf}
\end{figure}

\figref{perf} shows the speedups from two hardware extensions in \name{}:
hardware-based early termination (HET) and quad merging (QM). 
By applying each component to the baseline graphics hardware, we evaluate four
variants: Baseline, QM, HET, and HET+QM. 
QM reduces the amount of blending work for ROPs by partially blending the
overlapping quads within shader cores. 
This improves rendering performance by up to {1.49$\times$} over the baseline.
HET, on the other hand, excludes the fragments associated with terminated
pixels before fragment shading. This effectively reduces the number of
fragments to be processed in the hardware pipeline, thereby mitigating
unnecessary computations and memory accesses for both shading and blending,
providing a {1.80$\times$} speedup on average.
Overall, by combining HET and QM, \name{} provides an average speedup of
{2.07$\times$} over the baseline GPU.

The performance gain varies across the scenes, as the benefits of early
termination and quad merging are influenced by the scene size and viewpoint. 
For large, real-world outdoor scenes such as Train and Truck, we can achieve
greater speedups with early termination since a relatively large number of
Gaussians exist beyond the surface.
In contrast, Bonsai exhibits a lower speedup among real-world scenes due to its
inherent scene structure. With the object (i.e., the bonsai) positioned at the
center and surrounded by a background room, the benefit of early termination is
naturally concentrated in the central pixel region. 

For high-resolution scenes like Bonsai and Kitchen, the performance improvement
from quad merging is slightly lower compared to other scenes. This is because
successive primitives span a larger number of tile grids, causing TGC bins to
be flushed more frequently before being fully utilized.
Note that in practice, the primitives of these high-resolution scenes are
distributed across GPCs, which reduces the frequency of bin flushing.
Thus, the benefit of quad merging would likely be higher than observed in our
experiment.
In summary, \name{} consistently demonstrates performance improvements over the
baseline GPU, highlighting its broad applicability across diverse scene types.

\begin{figure}[t]
  \centering
  \includegraphics[width=\columnwidth]{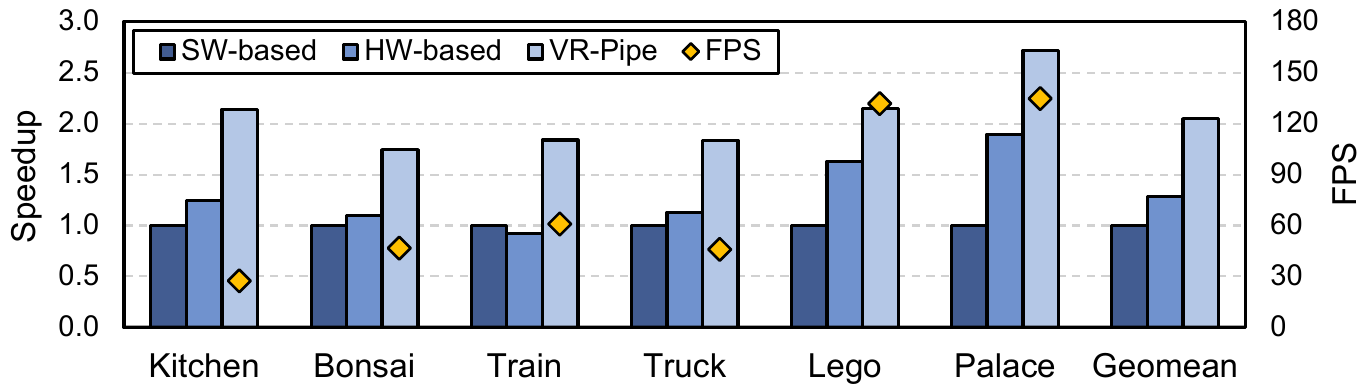}
  \caption{Overall end-to-end speedups of \name{} over software-based (CUDA) and
  hardware-based (OpenGL) rendering.}
  \label{fig:eval-perf}
\end{figure}

\myparagraph{Overall End-to-End Rendering.}
\figref{eval-perf} compares the overall end-to-end performance estimation,
which includes preprocessing and sorting, between different scenarios of 3D
Gaussian rendering.\footnote{As our evaluated GPU is configured similarly to
AGX Orin, we observe a high correlation in execution time between the AGX Orin
and our simulation for the preprocessing kernel. However, due to the limitations
of Emerald, it is challenging to obtain correlated execution time for the
sorting kernel that employs the NVIDIA CUB library.
To better estimate the overall rendering performance, we use the execution time
of preprocessing and sorting kernels on AGX Orin in this experiment.}
For a fair comparison, software-based (CUDA) rendering uses early termination,
while hardware-based (OpenGL) rendering does not, as the baseline architecture
lacks native support for early termination.
On average, \name{} can provide a 2.05$\times$ and 1.60$\times$ speedup
over the SW-based (CUDA) and HW-based (OpenGL) rendering, respectively. 
Since the rasterization step, which is our optimization target, occupies more
than 70\% of total rendering time, \name{} can effectively accelerate the
overall end-to-end rendering process, achieving high frame rates (FPS).


\begin{figure}[t]
  \centering
  \includegraphics[width=\columnwidth]{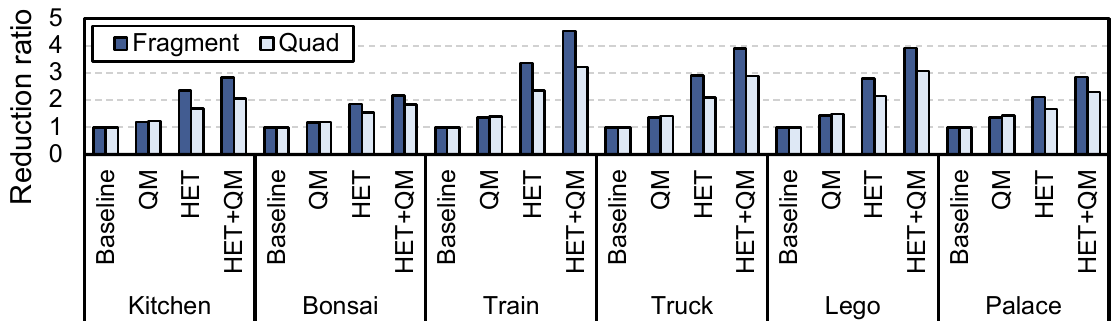}
  \caption{Reduction ratio of quads and fragments blended by ROP.}
  \vspace{-0.13in}
  \label{fig:frag-reduction}
\end{figure}

\putssec{src-of-gain}{Source of Performance Gain}

\figref{frag-reduction} shows the reduction in quads and fragments blended by
ROPs, a key source of performance gains for both HET and QM.
HET reduces the number of fragments by 2.52$\times$, significantly lowering the
amount of work by eliminating fragments before shading. QM further reduces the
number of fragments by 1.30$\times$ on top of HET, increasing the overall
blending throughput by utilizing shader cores for blending.
For quads, HET achieves a 1.90$\times$ reduction, with QM providing an
additional 1.32$\times$ reduction. The quad reduction of HET is smaller than
its fragment reduction because quads are eliminated only when all their
fragments are terminated. 
QM effectively reduces the number of quads by merging two into one. 
Consequently, by reducing ROP pressure, \name{} achieves speedups in relation
to the reduction in quads and fragments across the evaluated scenes.


\begin{figure}[t]
  \centering
  \includegraphics[width=\columnwidth]{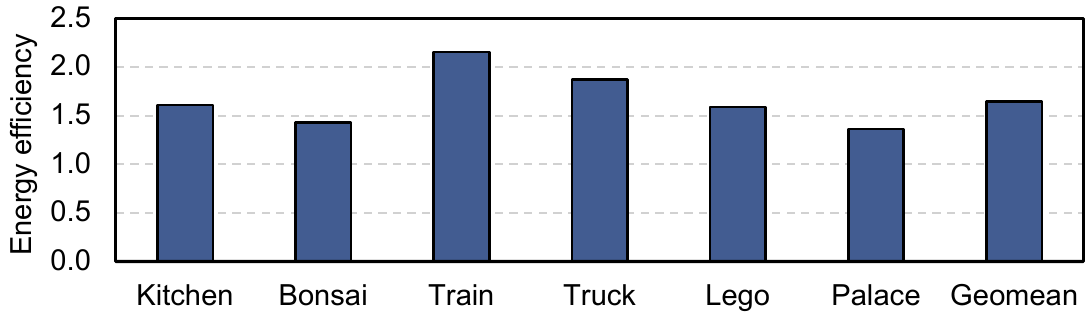}
  \caption{Energy efficiency of \name{} over the baseline GPU.}
  \vspace{-0.10in}
  \label{fig:energy-efficiency}
\end{figure}

\putssec{energy}{Energy Efficiency}
To estimate the energy efficiency of a draw call when \name{} is implemented
in mobile/embedded GPUs, we imitate the effect of HET and QM in the AGX Orin. 
For HET, we first count the number of fragments until early termination for each
pixel and store the numbers in the stencil buffer. 
During rendering, we then decrease the stencil value by 1 for each fragment
until the value reaches zero. Once the value becomes zero, we discard subsequent
fragments.
We also manually add warp shuffling, blending, discard operations in the
fragment shader, and make them operate as the same reduction ratio as QM in our
simulator. 
As a result, our approach provides {1.65$\times$} higher energy efficiency on average (up
to {2.15$\times$}) compared to the baseline GPU, as shown in \figref{energy-efficiency}.

\putssec{cost}{Implementation Cost}

\begin{table}[b]
  \caption{Hardware cost of \name{}.}
  \centering
    \resizebox{0.95\columnwidth}{!}{%
      \begin{tabular}{cc}
        \toprule
        \textbf{Hardware}             & \textbf{Size}   \\
        \midrule
        \multirow{2}{*}{Tile Grid Coalescing Unit}
          & (4B CBE pointer * 3 vertices * 16 entries   \\
          & + 2B tile grid ID) * 128 bins = 24.25KB     \\
        \midrule
        \multirow{2}{*}{Quad Reorder Unit}   
          & (4B CBE pointer + 6-bit quad pos.) * 128    \\
          & + 64 * 1B register + 16B bitmap = 688B      \\
        \midrule
        \textbf{Total}              & 24.92KB           \\
        \bottomrule
      \end{tabular}
    }
  \label{tab:hw-cost}
\end{table}

\tabref{hw-cost} shows the hardware cost of our proposed hardware extensions
for volume rendering. We do not account for the cost of computational units
because the bitwise operators used for alpha and early termination tests, the
comparators in the tile grid coalescing unit, and the two floating point
comparators in the alpha test unit would be considerably cheaper than the
storage overhead.
Our hardware extensions require only 25KB of storage, which is negligible
considering it is for each GPC, not for a single SM.

\putsec{analysis}{Analysis and Discussion}

\putssec{reverse}{Analysis on Real Graphics Hardware}

Emerald~\cite{gub:aam19} does not implement, or approximately models, the
graphics hardware units that we need to build upon (e.g., ROP, tile binning
hardware). To address this, we created OpenGL-based microbenchmarks to explore
the characteristics of these special-purpose units in real graphics hardware
and properly model them in our simulation framework.
The microbenchmarks render rectangles or triangles by adjusting various
parameters, including positions, color formats, the number of involved screen
tiles and rectangle overlaps. 
They are carefully designed and configured to stress the \emph{unit of
interest} while minimizing the impact on or from other units. 
For example, we restrict our analysis to a single GPC by positioning primitives
within the screen tiles mapped to a specific GPC, as each GPC contains its own
special-purpose units.
Also, as modern GPUs employ color compression, we attempt to bypass it by
generating different colors for each pixel via hashing.
We run them on NVIDIA Ampere GPUs and measure the data using Nsight Graphics.
OpenGL extensions including \texttt{\small GL\_NV\_shader\_thread\_group} and
\texttt{\small GL\_NV\_shader\_thread\_shuffle} are also used to collect data
from running warps formed through the graphics pipeline.

\begin{figure}[t]
  \centering
  \includegraphics[width=\columnwidth]{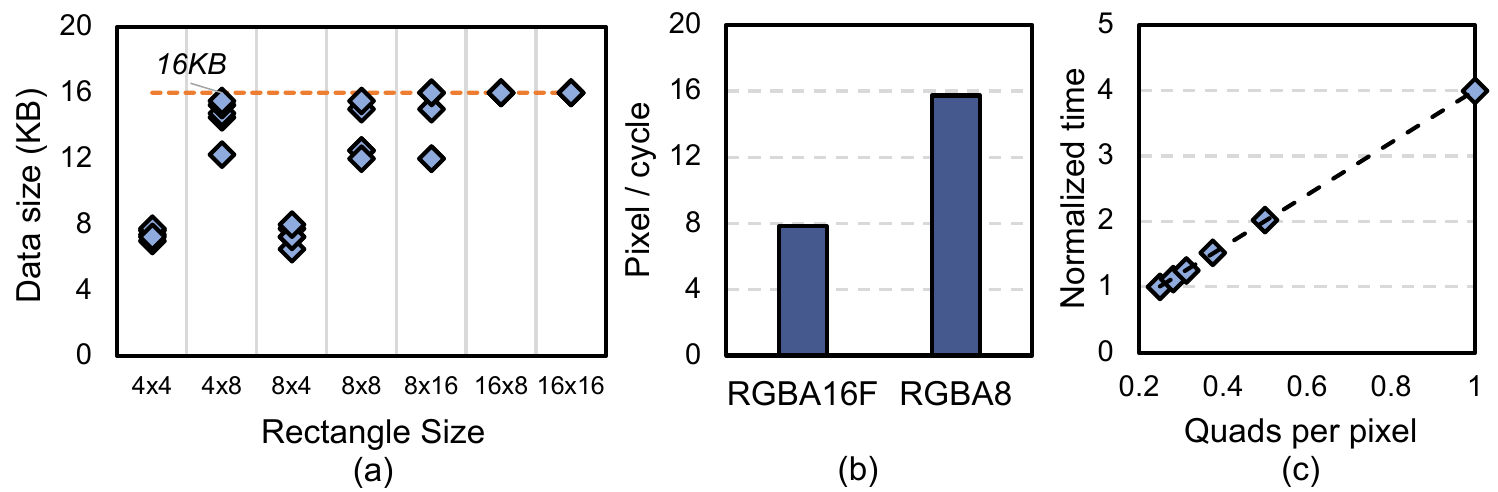}
  \caption{(a) Data sizes fitting within the CROP cache across different
  rectangle sizes, measured by rendering rectangles at randomized positions. The
  scatter in data points shows variability in cache usage due to randomized
  positions tested.
  (b) Pixels per cycle when rendering the same amount of pixels using
  RGBA16F or RGBA8 color format.  
  (c) Normalized rendering time with varying quads per pixel when using RGBA16F
  color format. The quads per pixel value is controlled by using a stencil test
  and adjusting the shape and position of primitives.
  }
  \vspace{-0.10in}
  \label{fig:rop-analysis}
\end{figure}

\myparagraph{Render Output Unit (ROP).}
The ROP units are now part of the GPC from the Ampere architecture to increase
the performance of raster operations~\cite{ampere}.
A single GPC has two ROP partitions, each containing eight ROP units. A single
ROP unit can process a single fragment by fetching the corresponding pixel
color of the image.
We observe that ROP units do not fetch pixel colors directly from the
L2 cache but first access another cache-like structure in each GPC, and we
refer to it as a CROP cache.
If a cache miss occurs, the color values are fed from the L2 cache. 

We first measure the size of the CROP cache by rendering multiple rectangles
while inspecting the L2 cache bandwidth consumption from CROP.
\figref{rop-analysis}(a) shows the size of pixel color data before CROP starts
accessing the L2 cache while increasing the number of rectangles. 
For example, we first draw an 8{$\times$}16-pixel rectangle using the
RGBA16F\footnote{Each red, green, blue, and alpha channel is a $16$-bit
floating-point number.} (FP16) pixel format, which corresponds to 1{KB}.
We then add more rectangles in random positions until the CROP starts accessing
the L2 cache.
The results show that 8{$\times$}16-pixel rectangles can be drawn at 16
locations without L2 cache access.
Over various rectangle sizes and positions, the CROP cache has never held more
than 16KB of data, suggesting its size is likely 16KB. 

\figref{rop-analysis}(b) shows that the number of pixels per cycle that CROP
can process depends on the color format of the rendered image. For example, the
pixels-per-cycle throughput is higher when using the RGBA8 (UNORM8) pixel
format compared to RGBA16F. For RGBA8, which is 32 bits per pixel (bpp), a GPC
with 16 ROP units can process 16 pixels per cycle, whereas it can process only
8 pixels per cycle when using the RGBA16F format, which is 64 bpp. 
This implies that the effective read bandwidth of the CROP cache is likely to
be larger than 64 bytes per cycle.
We also observe that ROPs cannot achieve the maximum pixels-per-cycle
throughput if some pixels in a quad are partially discarded, as shown in
\figref{rop-analysis}(c). This implies that the ROP units operate at a quad
granularity; i.e., four ROP units operate together to process a
2$\times$2-fragment quad.

\myparagraph{Tile Binning.}
We analyze the behavior of tile binning while varying the number of screen
tiles ($N$) involved in rendering.
To investigate the number of tile bins, we draw \emph{distinct} primitives
(2$\times$2-pixel rectangles) that are distributed across $N$ screen tiles and
measure the number of warps launched.
We arrange the rectangles to visit the screen tiles in a round-robin manner
(i.e., a repeating sequence of 1, 2, ..., $N$, 1, 2, ...), which helps us
clearly observe the flushing effect.
Note that each bin collects only the quads within the same screen tile, as
previously discussed.

We observe that quads from distinct rectangles at the same pixel position but
from different rounds are binned and launched together mostly as a single warp
if $N$ is less than or equal to 32. 
However, once $N$ exceeds 32, rectangles within the same screen tile but from
different rounds are launched as separate warps. 
For example, drawing 320 rectangles across 32 screen tiles results in 67 warps
being launched.
In contrast, drawing 330 rectangles across 33 screen tiles leads to the launch
of 330 warps (i.e., each warp contains only a single quad).
This occurs because, after the 32nd primitive, the binning of the 33rd
primitive---rendered on the 33rd screen tile---triggers the flushing of one of
the bins, resulting in the launch of an underutilized warp. This pattern
continues for the remaining primitives, which indicates that each GPC in the
evaluated GPUs has 32 tile bins.

\putssec{}{Early Termination Ratio of Varying Viewpoints}

\begin{figure}[t]
  \centering
  \includegraphics[width=0.95\columnwidth]{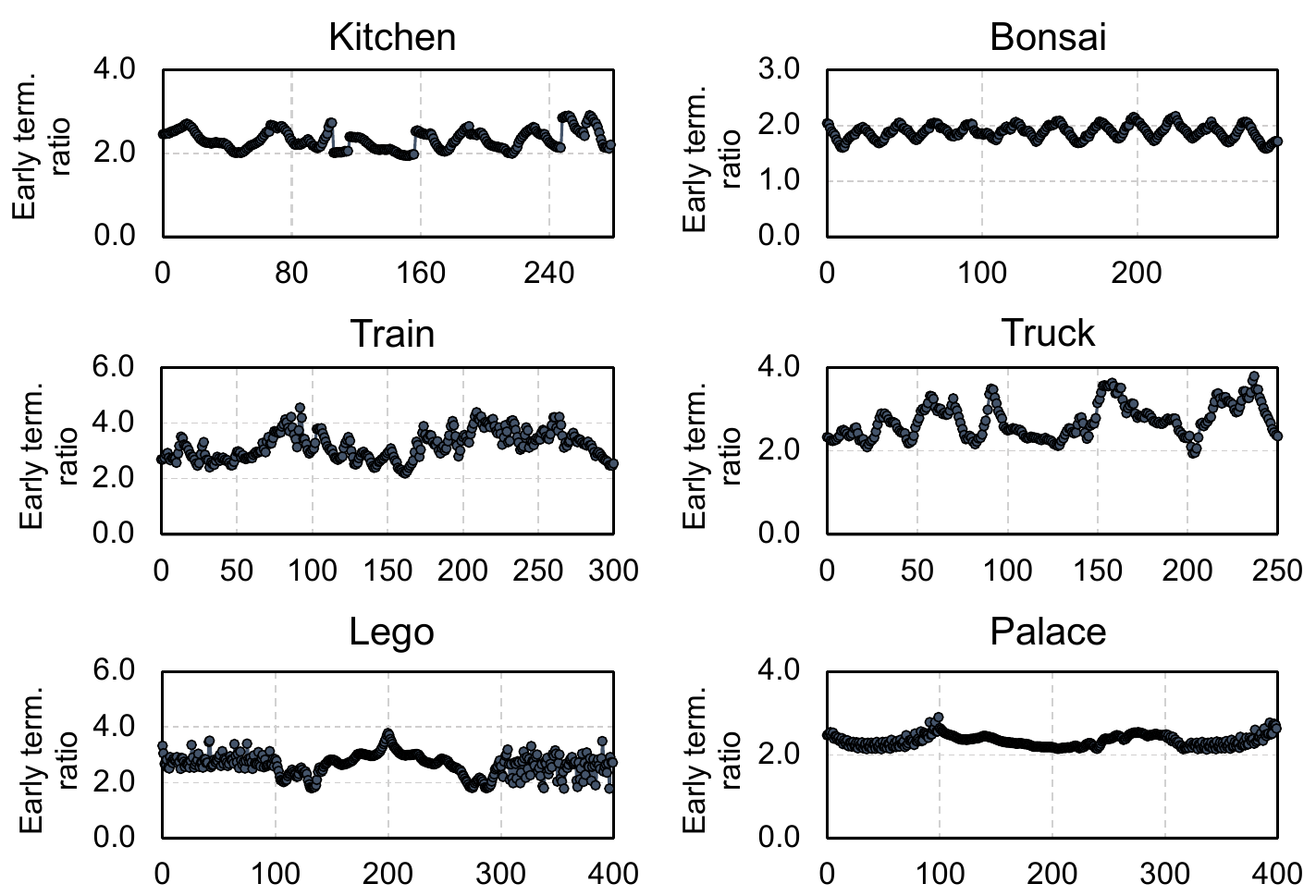}
  \caption{Early termination ratio of varying viewpoints.}
  \vspace{-0.10in}
  \label{fig:ert-ratio}
\end{figure}

\figref{ert-ratio} shows the early termination ratio across different
viewpoints. We evaluate all the viewpoints that are provided in the dataset.
The early termination ratio is the ratio between the number of blended
fragments processed with and without early termination. A higher ratio
indicates a greater potential for performance improvements through early
termination. 
As discussed in~\ssecref{src-of-gain}, outdoor scenes exhibit higher average
ratios than indoor or synthetic scenes because there are more Gaussians beyond
the surface in larger scenes. For example, up to nearly 4.4 times more
fragments could be unnecessarily blended without early termination in the Train
scene.
While the ratio varies depending on the scenes, every scene shows an average
ratio greater than 1.5, indicating that more than 33\% of the fragments can be
eliminated by early termination. 
This demonstrates that supporting early termination in hardware is a key factor
for efficient volume rendering.

\begin{figure}[t]
  \centering
  \includegraphics[width=\columnwidth]{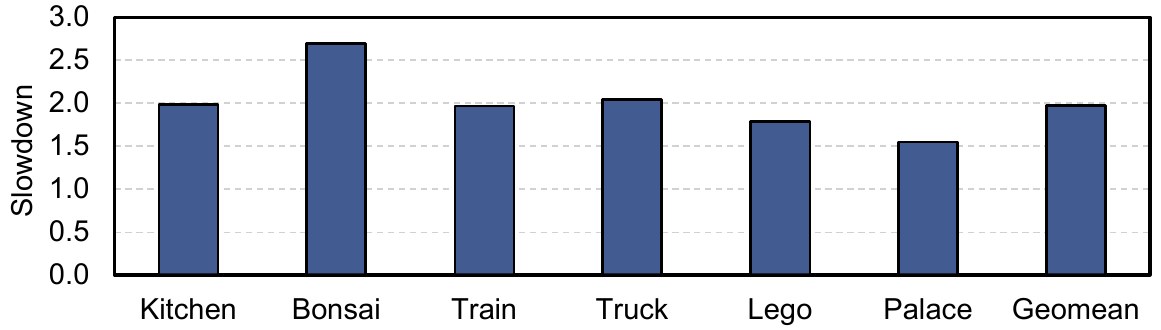}
  \caption{Performance comparison with GSCore~\cite{lee:lee24}.}
  \vspace{-0.1in}
  \label{fig:vr-pipe-vs-gscore}
\end{figure}

\begin{figure}[t]
  \centering
  \includegraphics[width=\columnwidth]{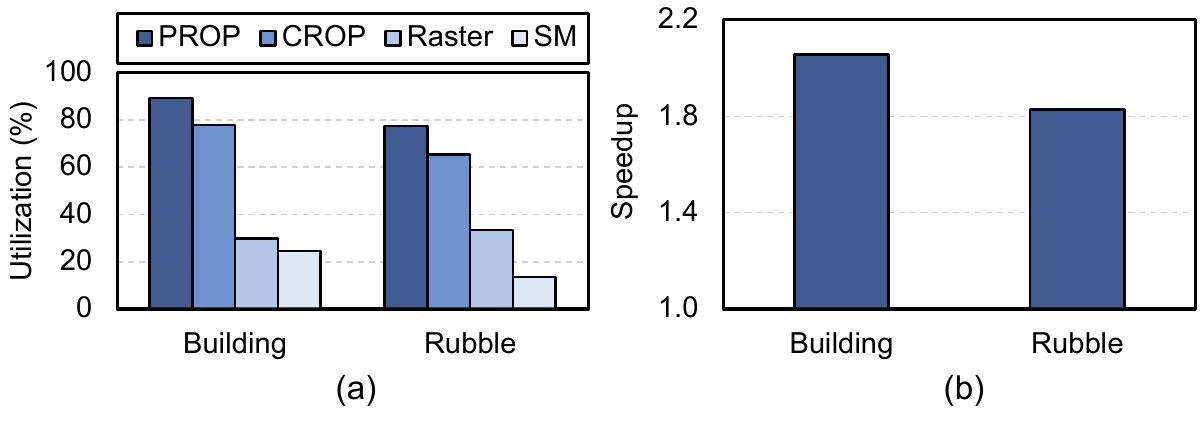}
  \caption{(a) Unit utilization and (b) speedup for large-scale scenes.}
  \vspace{-0.20in}
  \label{fig:large-scale}
\end{figure}

\putssec{}{Discussions}

\noindent\textbf{Comparison to Accelerators for 3D Gaussian Splatting.}
\figref{vr-pipe-vs-gscore} compares the performance of \name{} with a
specialized accelerator for 3D Gaussian splatting, GSCore~\cite{lee:lee24}. 
The results show that GSCore performs better than \name{} due to its tailored
design for Gaussian splatting.
As a dedicated accelerator, however, GSCore is inherently limited to running
graphics workloads involving Gaussian splatting and requires custom compilers
and runtime. 
\name{}, on the other hand, extends the capabilities of existing graphics
hardware and runs standard graphics APIs, offering greater flexibility
for rendering both traditional rasterization and volume rendering tasks.
We expect that future acceleration efforts will involve both building
specialized accelerators and enhancing existing graphics hardware, depending on
deployment requirements.

\myparagraph{Scalability for Larger or More Complex Scenes.}
As mentioned in~\ssecref{gr}, Gaussian primitives are rendered in a tile-based
manner, and thus the benefit of \name{} can be easily extended to very
large-scale scenes as long as they fit within the GPU memory.
For larger or more complex scenes, such as Building (9.06M Gaussians) and
Rubble (5.21M Gaussians) used in Mega-NeRF~\cite{tur:ram22} and
CityGaussian~\cite{liu:gua24}, there is a significantly higher number of
Gaussians that need blending, which makes ROPs a persistent bottleneck, as
shown in~\figref{large-scale}(a). Consequently, \name{} helps improve rendering
performance for these large-scale scenes, as shown in~\figref{large-scale}(b).

\myparagraph{Potential for Future Deployments in GPUs.}
Both hardware early termination (HET) and quad merging (QM) could potentially
be adopted in GPUs, as the main hardware overhead is only tens of kilobytes of
storage for the bins in the TGC unit and a few registers in the quad reorder
unit, as discussed in~\ssecref{cost}.
Between HET and QM, we envision that HET may have great potential for direct
deployment, as it offers solid performance benefits across different scene
types, as shown in~\figref{perf}, and is relatively simple to implement without
much hardware overhead.
As for QM, it is noted that the storage overhead added for QM is per GPC, not
per SM. Given that the baseline GPU (Jetson AGX Orin) contains at least 3.6MB
of SRAM in a single GPC, the additional overhead for QM might also be
acceptable.

\putsec{related}{Related Work}

\noindent \textbf{Efficient Radiance Field Rendering.}
The introduction of Neural Radiance Fields (NeRF)~\cite{mil:sri20} has
generated significant interest in efficient 3D scene representation and
rendering for radiance fields.
Over the past years, there has been a large amount of research aimed at
accelerating NeRFs through algorithmic or software
optimizations~\cite{mul:eva22,fri:yu22,che:fun23,sun:sun22}, and the
development of hardware
accelerators~\cite{lee:cho23,li:li23,son:wen23,mub:kan23,fen:liu24}.
The state-of-the-art method, 3D Gaussian splatting~\cite{ker:kop23}, has
further fueled interest in accelerating radiance field
rendering~\cite{rad:ste24,lee:lee24,nie:stu24,lee:rho24,ham:mel24} as it
employs rasterization primitives that can be rendered much faster than NeRFs.
However, previous research focused on software graphics rendering on
programmable cores or building dedicated hardware accelerators. In contrast,
\name{} investigates the potential of efficient radiance field rendering while
utilizing fixed-function units in graphics hardware.
To our knowledge, this is the first work that assesses the performance
implications of rendering Gaussian-based radiance fields on the hardware
graphics pipeline with software and hardware optimizations.

\myparagraph{Enhancing Graphics Rendering Hardware.}
The performance advantage of executing graphics rendering on either
programmable shader cores or fixed-function units varies depending on the
rendering methods and hardware designs.
Previous studies have explored the performance implication of graphics hardware
design by developing simulation infrastructures for graphics
workloads~\cite{bar:gon06,gub:aam19,tin:sax23,arn:par13}.
Additionally, several studies have aimed to improve the performance of
special-purpose hardware such as ray tracing units in graphics
hardware~\cite{cho:now23,liu:cha21} and proposed hardware accelerators for
graphics applications~\cite{lu:hua17,ram:gri09}.
In contrast to these works, which primarily evaluate traditional graphics
workloads, our work focuses on improving the performance of volume rendering
workloads, such as Gaussian splatting, which require blending a huge number of
fragments per pixel.

%
In the context of multi-sample anti-aliasing, prior work proposed reducing the
amount of redundant shading by merging fragments from adjacent triangles in a
mesh at the quad granularity~\cite{fat:bou10}.
While both our work and quad-fragment merging (QFM)~\cite{fat:bou10} aim to
reduce operations by merging quads, our proposed technique differs from QFM in
many aspects.
Our method aims to blend \emph{overlapping primitives} along the depth
direction and applies to quads from any primitive. In contrast, QFM merges quad
fragments from small (e.g., pixel-sized) triangles that \emph{share} an edge
(i.e., \emph{connected}, \emph{non-overlapping} triangles).
As such, QFM is not applicable to the scenes consisting of a number of
unconnected transparent triangles, such as those in 3D Gaussian splatting.
In addition, our method computes the \emph{exact} color for each pixel by
offloading blending operations from ROPs to shader units, whereas QFM
\emph{approximates} pixel colors by using the color from one triangle when
multiple triangles are merged into a single quad.

\putsec{conclusion}{Conclusion}

Gaussian splatting is becoming one of the key techniques in graphics rendering,
yet its performance implication on the hardware graphics pipeline has remained
unexplored thus far.
Through hardware graphics rendering and software optimization using a standard
graphics API, we show that hardware-based volume rendering can match or surpass
the performance of software-based rendering that relies solely on shader cores.
Nonetheless, we also observe that there is potential for \emph{native} hardware
support for volume rendering workloads in graphics hardware.
Based on the analysis on modern graphics hardware, we introduce \name{}, which
integrates hardware-based early termination and multi-granular bins with quad
merging into the special-purpose units in GPUs. These features greatly
improve the performance of rendering scenes via volume rendering such as
Gaussian splatting. 
We anticipate that these features will also benefit graphics and game engines
that traditionally run on graphics-specific hardware.

\section*{Acknowledgment}
We thank the anonymous reviewers for their valuable feedback.
This work was supported in part by the Institute for Information \&
Communications Technology Planning \& Evaluation (IITP) grants funded by the
Korean government (MSIT) (RS-2023-00256081, RS-2024-00395134) and a
research grant from Samsung Electronics Co., Ltd. (0418-20230064, A0342-20200002).
The Institute of Engineering Research at Seoul National University provided
research facilities for this work. Jaewoong Sim is the corresponding author.


\bibliographystyle{IEEEtranS}
\bibliography{refs}

\end{document}